\documentclass{JFM-FLM_Au}

\usepackage{amssymb}
\usepackage{amsmath}
\usepackage{graphicx}
\usepackage{epstopdf,epsfig}
\usepackage{newtxtext}
\usepackage{newtxmath}
\usepackage{upgreek}
\usepackage{natbib}
\usepackage{float}
\usepackage{hyperref}
\usepackage{subcaption}
\usepackage{verbatim}
\hypersetup{
    colorlinks = true,
    urlcolor   = blue,
    citecolor  = black,
}

\newcommand{\RomanNumeralCaps}[1]
\linenumbers
\lefttitle{J. Proudfoot, C. J. Nicholls, B. M. T. Tang, and M. Bacic}

\righttitle{Journal of Fluid Mechanics}
\title{Feedback control of vortex shedding using data-driven modelling}

\author{Jack Proudfoot\aff{1},
\corresau{Marko Bacic, \email{marko.bacic@eng.ox.ac.uk}}
Chris J. Nicholls\aff{1},
Brian M.T. Tang\aff{1},
\and Marko Bacic\aff{1,2}}

\affiliation{\aff{1} Oxford Thermofluids Institute, University of Oxford, Oxford OX2 0ES, United Kingdom \aff{2}Rolls-Royce plc, Derby DE24 8BJ, United Kingdom}

\begin{document}
\maketitle

\begin{abstract}
This paper details the data-driven modelling and feedback control of vortex shedding past a circular cylinder at a Reynolds number of $\Rey$\,$=$\,$1000$. We study the effect of varying the order of the reduced model for control design purposes and demonstrate that higher orders can lead to lower suppression of vortex shedding. We use the Bode integral theorem and a frequency-domain interpretation to show that this drop in performance is, in part, due to the classical “waterbed effect,” which increases sensitivity in frequency bands of unmodelled dynamics. Training data from 2D unsteady simulation is used to obtain linear reduced-order state-space models of the system via dynamic mode decomposition with control. Using only lift measurement, we show that at least a $4th$-order model is required for an LQG controller to suppress vortex shedding, with the best performance achieved with as few as 9 modes, whilst higher-order (\textgreater14) controllers show a significant decrease in performance. We study the influence of external disturbances, noise rejection, and parameter uncertainty on controller performance. A $28.6$\,dB reduction in lift coefficient variance is achieved, resulting in a $26\%$ reduction in drag. We further show that, for control design purposes with practical actuation bandwidth, the closed-loop control delivers a significant $13.7\%$ drag reduction within 3D DDES, despite having been trained with 2D URANS and therefore argue that 2D URANS simulation is sufficient for reduced-order model generation and control design.
\end{abstract}

\begin{keywords}
Authors should not enter keywords on the manuscript, as these must be chosen by the author during the online submission process and will then be added during the typesetting process (see \href{https://www.cambridge.org/core/journals/journal-of-fluid-mechanics/information/list-of-keywords}{Keyword PDF} for the full list). Other classifications will be added at the same time.
\end{keywords}
\newpage
\section{Introduction}
\label{sec: introduction}
This paper studies the problem of feedback control of the simplest oscillator flow \citep{sipp_linear_2016}, that of cylinder vortex shedding, from a frequency domain perspective. For a smooth cylinder with low free stream turbulence intensity, a Hopf bifurcation results in laminar vortex shedding beyond a critical Reynolds number ($\Rey$ $\!\approx$\,$49$) \citep{Williamson_1996}, commonly called the Von Kármán vortex street, which is widely recognised as a non-linear limit cycle \citep{roussopoulos_1993}. Dynamics of such behaviour in the vicinity of the critical Reynolds number have been experimentally validated to follow the Stuart-Landau model \citep{Provansal_Mathis_Boyer_1987}. In practical terms, the onset of vortex shedding is associated with a rise in drag and unsteady lift forces, motivating researchers to attempt to eliminate, or at least suppress, wake unsteadiness \citep{henderson_1997}. As Reynolds number increases, shedding dynamics rapidly transition to a so-called `irregular regime', undergoing further bifurcations into spatio-temporal chaos \citep{roshko_1954}. Despite the inherent three-dimensionality of turbulent bluff-body flows, the primary instability triggered by the Hopf bifurcation is two-dimensional, dominating the energy spectrum and representing a coherent structure. Vortex shedding persists beyond a supercritical transition at $\Rey$\,$=$\,$3.5$\,$\times$\,$10^6$ \citep{roshko_1961}.
\par
Numerous studies attempting to control bluff-body shedding have been met with varying degrees of success, with a range of open-loop and feedback control approaches used. \citet{berger_1967} used a phase-shifted hot-wire for feedback in which vortex shedding is suppressed behind an oblong cylinder through transverse oscillations with small input amplitudes in low Reynolds number flows ($\Rey$\,$<$\,$300$). At higher input amplitudes, the excitation of higher-order modes has a destabilising effect, reducing the authority of the control strategy. Attempts to suppress vortex shedding using open-loop control include: blowing \citep{williams_1988}, suction \citep{gao_active_2019}, angular rotation \citep{tokumaru_1991}, wake heating \citep{schumm_1994}, and secondary cylinders in the near wake \citep{schulmeister_2017}. Open-loop studies provide insight into the dynamics of control but come with limitations, as they attempt to cancel dynamic behaviour using fundamentally uncertain implicit models of the real system. However, the primary purpose of feedback control is to reduce the uncertainty in the expected behaviour of the controlled plant in the presence of unmodelled dynamics and exogenous disturbances. Several researchers have therefore studied feedback control approaches \citep{monkewitz_1989, roussopoulos_1993, FUJISAWA_2001, Jin_2020}.
\par
\citet{monkewitz_1989} used a one-dimensional linearised Ginzburg-Landau model capable of partially suppressing a first-order vortex shedding mode with small control gains. Above a critical gain, no attenuation was achieved, which was attributed to the excitation of higher-order unmodelled instabilities. It was concluded that complete suppression of vortex shedding is implausible, and research should instead focus on achieving a reduction in global unsteadiness. Further studies using a single-input single-output (SISO) phase-shifted controller struggled to suppress shedding beyond the transitional Reynolds number experimentally \citep{roussopoulos_1993}. Similar experimental model-free approaches show some promise in the subcritical regime, with a phase lag controller optimally tuned using cylinder rotation \citep{FUJISAWA_2001}, but performance was highly sensitive to tuning parameters.
\par
Designing a high-performing, robust controller requires a well thought through reduced-order model (ROM). Analytical approaches utilising white-box models that rely on first principles \citep{monkewitz_1989, KrsticCylinder} can guide the control design to explore fundamentals of the problem, but they often, out of necessity, do not replicate the system's overall complexity. This is particularly true for fluid flow problems that can exhibit strongly non-linear dynamics, leading to the majority of recent approaches relying on grey box \citep{deane_1991, Jin_2020} or black box (data-driven) approaches \citep{Li_2024}.
\par
One successful grey box model reduction technique applied to the Navier-Stokes equations employs Galerkin models that extract the dominant dynamics of coherent structures through the projection of a reduced set of basis modes (avoiding the trap of dimensionality) \citep{Holmes_Lumley_Berkooz_1996}. \citet{deane_1991} used Galerkin models to describe unactuated cylinder flow accurately with relative simplicity. As few as eight basis modes were required to represent over $99.9\%$ of total energy at low Reynolds numbers. \citet{NOACK_2003} developed an improved Galerkin model by implementing a shift mode to capture transient evolutions from a symmetric unstable condition to the natural vortex shedding limit cycle. Such models have enabled feedback control with some success \citep{gerhard_2003}. However, when attempting to model actuated flows, the number of basis modes must increase to capture the added complexity of actuator-flow-field interactions. The introduction of actuator dynamics represents an inhomogeneous boundary condition, meaning the standard Galerkin expansion is no longer valid without the use of additional terms in that expansion \citep{graham_1999_1}. Such higher-order models have been used by \citet{graham_1999_2} in a numerical study of a rotating cylinder 
to achieve a reduction in wake unsteadiness through a penalty minimisation approach.
\par
An alternative to the Galerkin projection method is to apply resolvent analysis on the linearised Navier-Stokes equations and obtain an input-output transfer function \citep{Jin_2020}. Up to $\Rey$\,$=$\,$120$ the authors fit models of the order up to $N$\,$=$\,$30$ and apply {$\mathcal{H}_{\infty}$} design methodology. Two different actuator/sensing pairs are investigated, one with a transverse position of the cylinder coupled with direct lift measurement, and the other with momentum injection on the surface of the cylinder coupled with a velocity sensor on the centre-line.
\par
Notwithstanding the modelling structure or control methodology used, there are several additional challenges encountered by both \citet{graham_1999_2} and \citet{Jin_2020}. Clearly, as shown by \citet{graham_1999_2}, the choice of controlled variable is of critical importance, with the choice of wake unsteadiness acknowledged as a potential limiting factor to performance. Similarly, \citet{illingworth_2014} and \citet{Jin_2020} demonstrate the importance of transfer function zero locations on controller performance and resulting stability margin, particularly as those zeros become non-minimum phase \citep{Jin_2020}. This is not surprising, as it is well known that non-minimum phase zeros place fundamental limits on controller performance and robustness \citep{FreudenbergLooze, ASTROM20002}. Since the zeros depend on the location of sensors as well as underlying system dynamics, the choice of type and sensor locations should always be made, if possible, to result in input/output dynamics that are minimum phase.
\par
\subsection{Problem statement}
All of the feedback control studies mentioned above focus on $\Rey$\,$<$\,$150$, where the vortex street is laminar, with no study attempting to control increasingly turbulent vortex streets occurring at $\Rey$\,$>$\,$300$ using linear models. In this paper, we investigate SISO feedback control for the case of $\Rey$\,$=$\,$1000$, the beginning of the so-called ``shear layer transition regime", where three-dimensionality both on the scale of shear layer vortices and on the scale of Karman vortices are expected to develop \citep{Williamson_1996}. We consider here the objective of vortex suppression by proxy control of unsteady Kalman filter estimates of the lift coefficient ($C_\text{l}$). We utilise flow injection at slots in the cylinder surface, and apply dynamic mode decomposition with control (DMDc) methodology to identify our reduced-order linear state-space models. We use $\mathcal{H}_2$ Linear Quadratic Gaussian (LQG) methodology for control design as a convenient tool that can offer a degree of robustness subject to careful design \citep{doyle_lqg,doyle_stein_ltr}. We focus our efforts on answering the following research questions:
\begin{enumerate}
\item\, What is the effect of controller and model order on the magnitude of vortex suppression, and are there fundamental reasons why there might be an optimum model order?
\item\, What is the maximum model order that should be used for this case, given limited actuation authority?
\item\, How robust is the control and model to external disturbances, sensor noise, and model uncertainty due to unmodelled non-linear dynamics?
\item\, How effective is a control design methodology that utilises cheap 2D URANS data for training in a more representative 3D delayed detached eddy simulation (DDES) setup?
\item\, What is the effect of reduced actuation authority?
\end{enumerate}
\par
In \S\ref{sec: Method}, a detailed description of the computational set-up, modelling, and control algorithms is given. We validate the model with open-loop predictions of future state evolution subject to sinusoidal forcing. The influence of model rank is investigated, and the controller performance degradation of higher-order models is attributed to the waterbed effect. In \S\ref{sec: Results}, the results of the closed-loop control are presented for the LQG design with integral action for reference tracking. The robustness of the controller to sensor noise, external disturbances, and model uncertainty is presented. 
Finally, the robustness of the controller performance is evaluated in a 3D domain with DDES to assess the validity of controller designs based on a 2D training set.
\section{Methodology}\label{sec: Method}
\subsection{Problem domain}\label{sec: Domain}
We numerically investigate the flow past a circular cylinder with trailing-edge slots for actuation. All geometry is 2D unless otherwise stated. A Cartesian coordinate system is used with the origin centred on the cylinder. Where $x,y,z$ denote the streamwise, vertical, and spanwise directions, respectively, the domain is normalised with respect to the cylinder diameter $D$. The domain consists of a D-shaped velocity inlet ($\Gamma_1$) with radius $15D$ centred on the origin. Uniform inlet velocity is fixed to maintain a desired diameter-based Reynolds number, i.e. $U_{\infty}$\,$=$\,$(\Rey \mu)/(D \rho)$. A Dirichlet boundary condition at the outlet ($\Gamma_4$) maintains zero gauge pressure ($p$) $23D$ downstream of the origin. The top and bottom of the domain ($\Gamma_{2,3}$) are symmetry boundaries at $y$\,$=$\,$\pm 15D$, $0$\,$\leq$\,$x$\,$\leq$\,$ 23 D$.
The boundary conditions at the exterior of the domain are
\begin{eqnarray}
    (u_x,u_y)\,=\,(U_\infty,0) \quad \mathrm{on} \quad \Gamma_1,~~~ \\[2pt]
    \left( \frac{\partial u_x}{\partial y},u_y \right) \,=\,(0,0) \quad \mathrm{on} \quad \Gamma_{2,3}, \\[2pt]
    p\,=\,0 \quad \mathrm{on} \quad \Gamma_{4}. ~~
\end{eqnarray}
\par
The cylinder walls are smooth, no-slip, adiabatic walls. Actuation is achieved through slots ($\Gamma_\text{c}$) in the trailing edge of the cylinder. Two slots are controlled with a single desired actuation effort, labelled $\Gamma_\text{top}$ and $\Gamma_\text{bottom}$ to blow air into the wake depending on the sign of the input signal. The velocity at the actuation slot $u_\text{c}$ is determined by the control algorithm and updated at every time step.
\begin{eqnarray}\label{eq:control}
      \mathrm{if} \quad u_\text{c}(t)\geq 0, \quad \left\{
    \begin{array}{ll}
      u_{\text{top}}=u_\text{c}(t) \\[2pt]
      u_{\text{bottom}}=0
    \end{array} \right.
      \qquad ~~~~ \mathrm{else}, \quad  \left\{
    \begin{array}{ll}
      u_{\text{top}}=0 \\[2pt]
      u_{\text{bottom}}= \lvert u_\text{c}(t) \rvert
    \end{array} \right. \quad \mathrm{on} \quad \Gamma_\text{c}.
\end{eqnarray}
\par
\begin{figure}[h!]
\centerline{ 
\includegraphics[width=\textwidth]{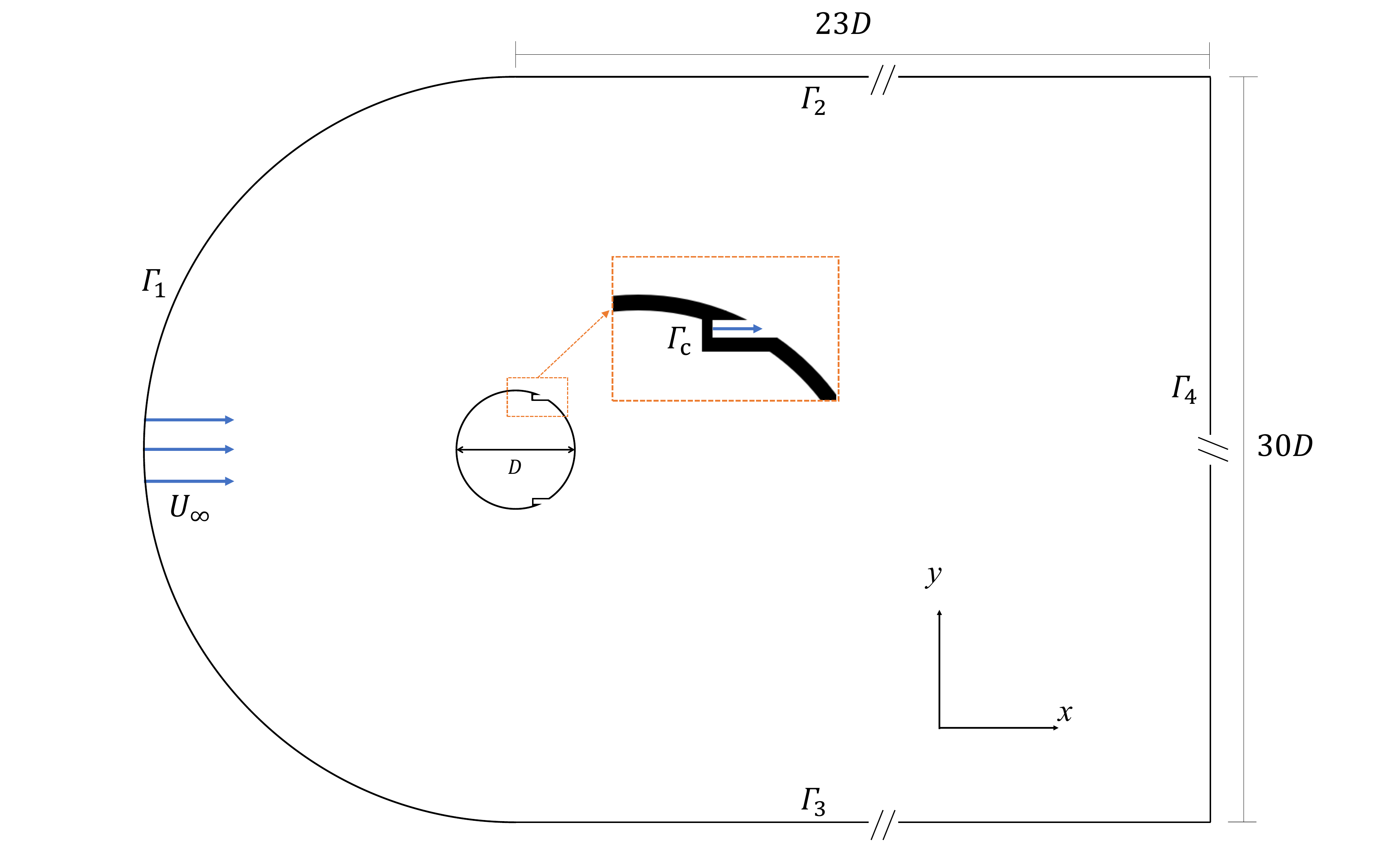}}
\caption{Geometry of computational domain (not to scale). Dimensions are normalised with respect to cylinder diameter $D$. The origin is at the centre of the cylinder, and Cartesian coordinates are used.}\label{fig: geom}
\end{figure}

The slots are located at $ \theta$\,$=$\,$\pm110^\circ$ from the leading-edge stagnation point. This ensures actuation is behind the separation point of the cylinder in the subcritical regime $\theta_\text{s}\approx 80^\circ$ \citep[p. 22]{schlichting_2017}. This actuation strategy is therefore not intended to manipulate the boundary layer and prevent separation, which likely requires a far higher bandwidth controller \citep{Glezer_2005}, but instead to interact with and break up the formation of trailing edge vortices in the wake to stabilise the flow field. The location of the actuation slots is similar to previous studies by \citet{park_1994} and \citet{illingworth_2014}. However, slots are physically modelled for our case rather than a direct modification of the surface boundary conditions, and suction is not used in our case. The height and setback distance of the slot are $5\%$ and $10\%$ of the cylinder diameter, respectively. The slot geometry is in part motivated by practical design limits, with the intent to build and test such a setup experimentally in a later study. The geometry is illustrated in figure \ref{fig: geom}. 
\subsection{Computational setup}\label{sec: setup}
To simulate flow past a cylinder, we solve the unsteady, incompressible Navier-Stokes equations
\begin{equation}
        \nabla \boldsymbol{\cdot u}=0,\quad   
    \frac{\partial\boldsymbol{u}}{\partial t}+(\boldsymbol{u\cdot} \nabla )\boldsymbol{u}=-\frac{1}{\rho}\nabla p+\nu \nabla^2\boldsymbol{u},
\end{equation}
using \textsc{Ansys Fluent}, where velocity $\boldsymbol{u}$ and pressure $p$ are functions of position and time, $\rho$ is the fluid density and $\nu$ the kinematic viscosity. The working fluid is air, and the Mach number (\textit{Ma}) is much less than 0.3; thus, the flow is considered incompressible ($\rho$\,$=$\,$1.225 \mathrm{kg}/\mathrm{m}^3$). The kinematic viscosity ($\nu$) is constant as the fluid is adiabatic and incompressible, $\nu$\,$=$\,$1.461\times10^{-5}\mathrm{m}^2/\mathrm{s}$. Despite the cylinder wall boundary layers being laminar, the vortices in the wake are turbulent beyond $\Rey$\,$>$\,$300$ \citep[p. 22]{schlichting_2017}. To model the effects of turbulence, the $\mathit{k}$\,-\,$\omega~ \mathrm{SST}$ model is used. It is well known that 2D URANS simulations of flow past a cylinder degrade in accuracy as the Reynolds number increases and are prone to overpredicting average drag \citep{karniadakis_1992, mittal_1995, travin_2000, elmiligui_2004}. Despite this overprediction of drag, shedding frequency ($f_\text{s}$) predictions are generally acceptable. Whilst the reduced-order model and subsequent controller are based on results from a 2D URANS formulation, we test the effectiveness of the controller against a 3D DDES \citep{Spalart_2001}. The results for the 3D DDES simulation are presented in \S\ref{sec: SRS}.
\par
For the training of models, a coupled, implicit, density-based formulation is used. Spatial and temporal discretisation is second-order accurate, and gradients are evaluated using a least squares cell-based formulation. The time step $\upDelta t$ is set to ensure that $\mathrm{CFL}$\,$\leq$\,$1$ in the cylinder wake, resulting in 775 simulation steps per shedding cycle. The controller sample interval is equal to the simulation time step. Time is non-dimensionalised with respect to the shedding frequency for the baseline configuration with no control.
\begin{equation}
    \tau=tf_\text{s}
\end{equation}
\par
The present results are compared against previous experimental and computational studies by comparing average drag ($\bar{C}_\text{d}$) and Strouhal number ($St$), and average base pressure coefficient ($\bar{C}_{p,\text{b}}$). Aerodynamic forces are normalised by free stream dynamic pressure
\begin{equation}
    C_\text{l}=\frac{2F_\text{l}}{\rho U_\infty^2 D},
\end{equation}
\begin{equation}
    C_\text{d}=\frac{2F_\text{d}}{\rho U_\infty^2 D},
\end{equation}
\begin{equation}
    C_{p,\text{b}}=\frac{2(p_\text{b}-p_\infty)}{\rho U_\infty^2},
\end{equation}
\begin{equation}
   St=\frac{f_\text{s}D}{U_\infty},
\end{equation}
where $F_\text{l}$ and $F_\text{d}$ are the integrals of vertical and horizontal forces acting on the cylinder walls, respectively. The net force due to actuation is purely horizontal and can be ignored for lift calculations. The momentum flux of the actuators ($\Dot{m}u_\text{c}$) produces a positive horizontal force, which is removed from the drag coefficient calculation to ensure only the aerodynamic influence of the control strategy is considered when assessing performance, i.e. $F_\text{d}$\,$=$\,$F_x$\,$-$\,$\dot{m}u_\text{c}$. 
The shedding frequency, $f_\text{s}$, is found by applying Welch's method \citep{Welch_67} to the lift force signal for at least ten oscillations in the limit cycle. 
\par
The domain is discretised into an unstructured grid with local refinement for boundary layers and wake dynamics. The near-wall cell height is set to maintain a wall-adjacent cell $\mathit{y^+}$\,$\leq$\,$1$ along the cylinder walls, with at least 16 cells in the boundary layer. The boundary layer thickness is estimated using the relationship $\delta_1$\,$\sim$\,$\textit{O}(D/\sqrt{\Rey})$ \citep[pp. 206-207]{schlichting_2017}. The total inflation layer height is subsequently adjusted to resolve the boundary layer completely.
\par
The grid is successively refined to confirm grid independence using the grid convergence index (GCI) \citep{roache_1994}. Three grids are tested with effective refinement ratios of
\begin{eqnarray}
    r_{21}=\left(\frac{119367}{54656}\right)^{1/2}=1.478 \\
    r_{32}=\left(\frac{54656}{26799}\right)^{1/2}=1.428
\end{eqnarray} which are approximated to be $\sqrt{2}$. A summary of grid refinement results compared to previous experimental and computational work is presented in table \ref{table:grid}.
\begin{table}
  \begin{center}
\def~{\hphantom{0}}
  \begin{tabular}{lcccc}
  Grid No. & Cell Count & St & $\bar{C}_\text{d}$ & $-\bar{C}_{p,\text{b}}$\\
       3   & 26799 & ~~0.21~ & 1.2397 & - \\
       2   & 54656 & ~~0.21~ & 1.1938 & 1.35\\
       1  & 119376 & ~~0.21~ & 1.1701 & - \\
       \citet{Rosetti_2012} & - & ~~0.23~ & 1.45 & 1.51\\
       \citet{braza_1986} & 31476 & ~0.21 & 1.19 & - \\
    \citet{henderson_1997} & - & ~~0.21~ & 1.20 & 1.01 \\
    \citet{Norberg_1987} & - & ~~0.21~ & 0.98 & 0.84
  \end{tabular}
  \caption{Summary of grid refinement for $\Rey$\,$=$\,$1000$, comparing to 2D results by \citet{braza_1986} and \citet{Rosetti_2012}, 3D DNS results by \citet{henderson_1997} are at $\Rey$\,$=$\,$1000$, and 3D experimental results by \citet{Norberg_1987} are at $\Rey$\,$=$\,$3$\,$\times$\,$10^3$.}\label{table:grid}
  \end{center}
\end{table}
The difference ratio for the average drag coefficient between successive grids is 1.937, resulting in an estimated convergence rate of
\begin{equation}
    p =\ln(\epsilon_{32}/\epsilon_{21})/\ln(\sqrt2)\approx1.9072.
\end{equation}
Using Richardson extrapolation, we may estimate the value of the drag coefficient with zero grid spacing
\begin{equation}
    P_r=\bar{C}_{\text{d},1}+(\bar{C}_{\text{d},1}-\bar{C}_{\text{d},2})/(\sqrt{2}^p-1)=1.1448.
\end{equation}
To estimate convergence, we use the grid convergence index (GCI) given as
\begin{eqnarray}
    GCI_{21} = \frac{F_\text{S}(\bar{C}_{\text{d},1}-\bar{C}_{\text{d},2})}{\bar{C}_{\text{d},1}(\sqrt{2}^p-1)}\times100=2.7\%, \\
    GCI_{32} = \frac{F_\text{S}(\bar{C}_{\text{d},2}-\bar{C}_{\text{d},3})}{\bar{C}_{\text{d},2}(\sqrt{2}^p-1)}\times100=5.1\%, \\          
\end{eqnarray}
where $F_\text{S}=1.25$ is a factor of safety. Finally, we can check that the solutions are within the asymptotic range of convergence
\begin{equation}
    GCI_{32}/(\sqrt{2}^pGCI_{21})=0.98.
\end{equation}
\par
We can therefore conclude that the grids are within the asymptotic convergence range and the grid is suitably insensitive to spacing.
$\mathrm{GCI}_{21}$ is below $5\%$ approaching the asymptotic solution; thus, grid 2 is used for the remaining study.
\par
$St$\,$=$\,$0.21$ compares favourably to previous studies by \citet{roshko_1954}. The average drag coefficient, $\bar{C}_\text{d}$\,$=$\,$1.19$ agrees with previous 2D computational results by \citet{braza_1986}, which acknowledges the limited capability of a two-dimensional simulation to capture inherently three-dimensional dynamics due to secondary instabilities in the wake \citep{henderson_1997}. Comparisons to experimental results show significant deviation in drag and base pressure predictions, as expected.
\subsection{Modelling and control design}\label{sec: model}
To create a suitable state-space model for control design, we utilise the dynamic mode decomposition with control (DMDc) algorithm \citep{Rowley_2009}. Dynamic mode decomposition (DMD) was first introduced as a data-driven regression method to extract a reduced-order linear system using experimental or simulated snapshots viewed as a truncation of the infinite-dimensional linear Koopman operator \citep{Rowley_2009, Schmid_2010} representing fundamentally finite-dimensional non-linear dynamics.
The resulting DMD model yields a discrete state-space autonomous model describing the dynamics of the reduced basis modes, which approximate the flow field
\begin{equation}
    \boldsymbol{x}_{k+1}\approx \mathsfbi{A}\boldsymbol{x}_k,
\end{equation}
where $\mathsfbi{A}$ is the state-transition matrix. Training snapshots consist of sequential stacked data vectors $\mathsfbi{X,~X'}$\,$\in$\,$\mathrm{\Real} ^{m \times (n-1)}$, where $m$ is the size of the discretised domain used for training, and $n$ is the total number of snapshots.
\begin{equation}
    \mathsfbi{X}=\begin{bmatrix}
    \mid & \mid & & \mid \\
    \boldsymbol{x}_1 & \boldsymbol{x}_2 & \dots & \boldsymbol{x}_{n-1} \\
    \mid & \mid & & \mid \\
    \end{bmatrix},    \quad \mathsfbi{X'}=\begin{bmatrix}
    \mid & \mid & & \mid \\
    \boldsymbol{x}_2 & \boldsymbol{x}_3 & \dots & \boldsymbol{x}_{n} \\
    \mid & \mid & & \mid \\
    \end{bmatrix}.
\end{equation}
$\mathsfbi{A}$ may be computed directly through the pseudo-inverse ($^\dagger$) \citep{Tu_2014}
\begin{equation}
\mathsfbi{A}=\mathsfbi{X'}\mathsfbi{X^\dagger},
\end{equation}
this is computationally infeasible for $(m$\,$\times$\,$(n$\,$-$\,$1))$\,$\gg$\,$1$. Thankfully, the high-dimensional system evolves according to a low-dimensional attractor and has a compact representation in a reduced basis. This reduced basis is achieved through the truncated singular value decomposition (SVD) of the snapshot matrices $\mathsfbi{X}$\,$\approx$\,$\mathsfbi{U}_r\boldsymbol{\upSigma}_r\mathsfbi{V}_r^*$, where $^*$ is the complex conjugate transpose and $r$ is the size of the truncation rank ordered by the size of the singular values $\sigma_r$. Assuming the first $r$ modes sufficiently represent the dominant dynamics of the high-order system, the resulting approximate state-transition matrix is given as
\begin{equation}\label{eq:Admd}
\mathsfbi{A}\approx\mathsfbi{U}_r^*\mathsfbi{X'}\mathsfbi{V}_r \boldsymbol{\upSigma}^{-1}_r.
\end{equation}
\subsubsection{Dynamic Mode Decomposition with Control}\label{sec: DMDc}
When actuation is present, the DMD algorithm may be modified to incorporate the interaction of actuation with free dynamics into the DMDc algorithm introduced by \citet{Proctor_2016}, resulting in a model with structure
\begin{equation}
\boldsymbol{{x}}_{k+1}=\mathsfbi{A}\boldsymbol{x}_k+\mathsfbi{B}u_k.
\end{equation}
\par
The input-state matrix is given as $\mathsfbi{B}$, and the discrete-time inputs are given as $u_k$. In terms of the aforementioned snapshot matrices, this is given as
\begin{equation}
    \mathsfbi{X'} =
\begin{bmatrix}
    \mathsfbi{A} & \mathsfbi{B}
\end{bmatrix}
\begin{bmatrix}
    \mathsfbi{X} \\
    \boldsymbol{\Upsilon} 
\end{bmatrix}
=
\mathsfbi{G}\boldsymbol{\upOmega},
\end{equation}
where $\boldsymbol{\Upsilon}$\,$\in$\,$\mathrm{\Real}^{(n-1)}$ is the actuation input vector. Subsequently, the state-transition and input-state matrices may be approximated through the same process of truncation and projection in equation \ref{eq:Admd}
\begin{eqnarray}
\boldsymbol{\upOmega}\approx \mathsfbi{\hat{U}}_r\boldsymbol{\hat{\upSigma}}_r\mathsfbi{\hat{V}}_r^*, \\[2pt] \mathsfbi{X'}\approx \mathsfbi{U}_r\boldsymbol{\upSigma}_r\mathsfbi{V}_r^*,
\end{eqnarray}
where $\mathsfbi{A}$ is given by
\begin{equation}
\mathsfbi{A}\approx\mathsfbi{U}_r^*\mathsfbi{X'}\hat{\mathsfbi{V}}_r \hat{\boldsymbol{\upSigma}}^{-1}_r\hat{\mathsfbi{U}}^*_{r,1}\mathsfbi{U}_r,
\end{equation}
and $\mathsfbi{B}$ is given by
\begin{equation}
    \mathsfbi{B}\approx \mathsfbi{U}_r^*\mathsfbi{X'}\hat{\mathsfbi{V}}_r \hat{\boldsymbol{\upSigma}}^{-1}_r\hat{\mathsfbi{U}}^*_{r,2}.
\end{equation}
\par
We may project the eigenvectors of the state-transition matrix $\mathsfbi{A}$ by first solving the eigenvalue decomposition $\mathsfbi{A}\mathsfbi{W}\,$=$\,\mathsfbi{W}\boldsymbol{\upLambda}$ and producing the DMD modes $\boldsymbol{\upPhi}$ with
\begin{equation}
\boldsymbol{\upPhi}=\mathsfbi{X'}\hat{\mathsfbi{V}}_r \hat{\boldsymbol{\upSigma}}^{-1}_r\hat{\mathsfbi{U}}_{r,1}^*\mathsfbi{U}_r\mathsfbi{W},
\end{equation}
where each DMD mode is associated with a complex eigenvalue and is exactly the poles of the identified state-space model. DMDc generates models suitable for real-time control of high-dimensional non-linear systems through the assumption that significant spatio-temporal coherent structures evolve on a low-dimensional attractor \citep{Kutz_2016}. The use of some variant of DMD to generate ROMs for control is not new, with several studies implementing some variation of the standard algorithm \citep{noack_2016, deem_2020, Li_2024}. 
\subsubsection{Choice of observable}
\label{sec: observe}
To apply state feedback directly, we would need full state information, which in turn requires knowledge of the reduced flow field states $\boldsymbol{x}_k$. This would be impossible to deploy for a physical system. 
Instead, we introduce an open-loop mapping of the state vector to a measurable output variable ($y_k$). In this paper, we consider the case of direct measurements of the lift coefficient as our observed output. This leads to the introduction of an output-state matrix ($\mathsfbi{C}$\,),
\begin{equation}
C_\text{l}=y_k=\mathsfbi{C}\boldsymbol{x}_k,
\end{equation}
resulting in a SISO state-space realisation. $C_\text{l}$ is the preferred observable rather than the drag coefficient, $C_\text{d}$, because the minimum achievable drag coefficient, $C_\text{d,min}$, (stabilised flow) is not known a priori for a given problem, whilst it is self-evident that $C_\text{l}\,$=$\,0$ for a fully stabilised flow. 
\par
To produce the open-loop mapping, we use a least-squares regression fit of lift coefficient against state vector dynamics $\mathsfbi{C}\,$=$\,C_\text{l}^\text{T} \boldsymbol{x}^\dagger$, where $^\mathrm{T}$ is the matrix transpose. The accuracy of the open-loop mapping is directly correlated to the truncation rank. 

The open-loop mapping provides a good estimate of the unactuated lift coefficient and captures the dominant behaviour of the response to actuation. We do not require the model to be capable of predicting dynamics associated with high-amplitude, high-frequency inputs, as these would likely excite higher-order (unmodelled) modes with non-linear coupling. The subsequently identified model is a linear time-invariant SISO structure of size $r$ following
\begin{eqnarray}
\boldsymbol{{x}}_{k+1}=\mathsfbi{A}\boldsymbol{x}_k+\mathsfbi{B}u_k,\\
\boldsymbol{{y}}_{k}=\mathsfbi{C}\boldsymbol{x}_k .
\end{eqnarray}
\par
These resulting realisations are confirmed to be observable and controllable for all truncation ranks studied.
\subsubsection{Feedback control design}\label{sec: control}
Since the resulting DMD model structure is state-space, we utilise a classical Linear Quadratic Gaussian (LQG) \citep{Astrom_1996} methodology for control law design due to its convenient tuning framework. Starting from LQR controller design, we obtain a discrete state feedback law $u_k\,$=$\,-\mathsfbi{K}\boldsymbol{x}_k$ 
 by minimising the quadratic cost function
\begin{equation}
J=\sum_{k=0}^\infty (\boldsymbol{x}_k^\mathrm{T}\mathsfbi{Q}_\text{LQR}\boldsymbol{x}_k+\mathsfbi{R}_\text{LQR}u_k^2).\label{eq: costfnc}
\end{equation}
\par
The controller's closed-loop performance may be tuned by selecting weights of the positive definite matrices $\mathsfbi{Q}_\text{LQR}$ and $\mathsfbi{R}_\text{LQR}$ to penalise states and actuator input, respectively. The discrete-time infinite-horizon feedback gain is static and computed as
\begin{equation}
\mathsfbi{K}=\left(\mathsfbi{R}_\text{LQR}+\mathsfbi{B}^\mathrm{T}\bar{\mathsfbi{P}}\mathsfbi{B}\right)^{-1}\mathsfbi{B}^\mathrm{T}\mathsfbi{\bar{P}A},
\end{equation}
where the infinite horizon limit $\mathsfbi{\bar{P}}=\lim\limits_{k\to \infty}\mathsfbi{P}_k$ is the positive semidefinite solution to the discrete algebraic Riccati equation
\begin{equation}
\mathsfbi{A}^\mathrm{T}\mathsfbi{\bar{P}}\mathsfbi{A}+\mathsfbi{Q}_\text{LQR}-\mathsfbi{A}^\mathrm{T}\mathsfbi{\bar{P}}\mathsfbi{B}\left(\mathsfbi{R}_\text{LQR}+\mathsfbi{B}^\mathrm{T}\mathsfbi{\bar{P}}\mathsfbi{B}\right)^{-1}\mathsfbi{B}^\mathrm{T}\mathsfbi{\bar{P}}\mathsfbi{A}=\bar{\mathsfbi{P}}.
\end{equation} 
\par
Since we do not have access to full-state information, a Kalman filter is used to estimate the states ($\hat{x}_k$) and carry out LQG design. 
Similarly, for the infinite-horizon steady-state Kalman filter, its gain $\mathsfbi{L}$ is given by
\begin{equation}
    \mathsfbi{L}=\mathsfbi{A}\mathsfbi{\bar{P}}_\text{f}\mathsfbi{C}^\mathrm{T}(\mathsfbi{R}_\text{f}+\mathsfbi{C}\bar{\mathsfbi{P}}_\text{f}\mathsfbi{C}^\mathrm{T})^{-1},
\end{equation}
where the infinite horizon limit $\mathsfbi{\bar{P}}_\text{f}\,$=$\,\lim\limits_{k\to \infty}\mathsfbi{P}_{\text{f},k}$ is the positive semidefinite solution to the discrete algebraic Riccati equation
\begin{equation}
\mathsfbi{A}\mathsfbi{\bar{P}}_\text{f}\mathsfbi{A}^\mathrm{T}+\mathsfbi{Q}_\text{f}-\mathsfbi{A}\mathsfbi{\bar{P}}_\text{f}\mathsfbi{C}^\mathrm{T}(\mathsfbi{R}_\text{f}+\mathsfbi{C}\mathsfbi{\bar{P}}_\text{f}\mathsfbi{C}^\mathrm{T})^{-1}\mathsfbi{C}\mathsfbi{\bar{P}}_\text{f}\mathsfbi{A}^\mathrm{T}=\bar{\mathsfbi{P}}_\text{f}.
\end{equation}
\par
With the introduction of a Kalman filter, the closed-loop system no longer has automatic robustness guarantees \citep{doyle_lqg}, and loop-transfer recovery should be used to recover robustness \citep{doyle_stein_ltr, Maciejowski_1985} subject to systems being minimum-phase. However, in practice, the degree to which we may recover robustness is also a trade-off with sensor noise performance. Sources of noise include external disturbances from nominally uniform flow conditions, such as gusts and turbulent fluctuations.
\par
Whilst gusts consist of low frequencies and thus should be rejected by control action, turbulence is beyond the bandwidth of the controller and is treated here in the design as high-frequency sensor noise. The Kalman filter gains are therefore tuned to provide an acceptable trade-off, resulting in $\mathsfbi{Q}_\text{f}\,$=$\,1$, $\mathsfbi{R}_\text{f}\,$=$\,3\times 10^{-4}$. The subsequent controller's gain and phase margin are $27.7$\,$\text{dB}$ and $85.3^\circ$, respectively. The underlying assumption here is that even though the vortex shedding dynamics are non-linear, the large gain and phase margins are sufficient to keep the system stable
\begin{figure}[h!]
\centerline{ 
\includegraphics[width=\textwidth]{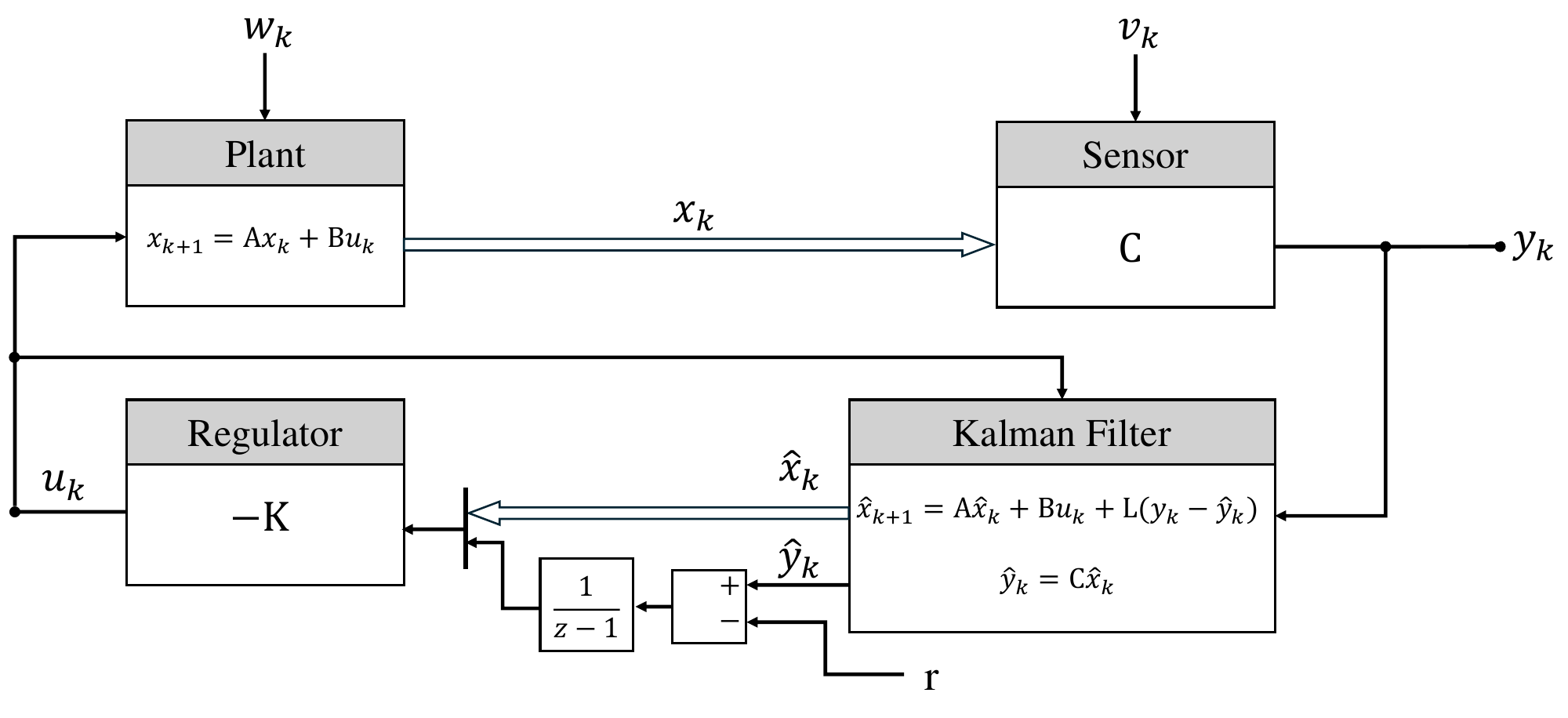}}
\caption{Linear Quadratic Gaussian block diagram with integral action and reference tracking.}\label{fig: model}
\end{figure}
\subsubsection{Integral Action}\label{sec: ref}
An extension to this control scheme is made by the addition of integral action, which enables reference tracking. The integrator ensures zero steady-state error by augmenting the state vector with an additional integral term
\begin{equation}
    x_{\text{i},k+1}=x_{\text{i},k}+r_k-\hat{y}_{k},
\end{equation}
where $\hat{y}_k$ is the Kalman estimate of the output observable and $r_k$ is the reference at discrete time $k$. This results in an augmented state vector
\begin{equation}
\boldsymbol{\tilde{x}}_{k} = \begin{bmatrix} \boldsymbol{x}_k \\ x_{\text{i},k} \end{bmatrix},
\end{equation}
which evolves according to the dynamics with additional process ($\mathsfbi{w}_k$) and sensor noise terms ($\mathsfbi{v}_k$)
\begin{eqnarray}
\begin{bmatrix}
\boldsymbol{x}_{k+1} \\
x_{\text{i},k+1}
\end{bmatrix}
=
\underbrace{
\begin{bmatrix}
\mathsfbi{A} & \mathsfb{0} \\
\mathsfbi{-C} & 1
\end{bmatrix}
}_{\tilde{\mathsfbi{A}}}
\begin{bmatrix}
\boldsymbol{x}_k \\
x_{\text{i},k}
\end{bmatrix}
+
\underbrace{
\begin{bmatrix}
\mathsfbi{B} \\
\mathsfb{0}
\end{bmatrix}
}_{\tilde{\mathsfbi{B}}} u_k
+
\begin{bmatrix}
\mathsfb{0} \\
\mathsfbi{I}
\end{bmatrix}
r_k
+
\begin{bmatrix}
\mathsfbi{w}_k \\
\mathsfb{0}
\end{bmatrix}, \\[2pt]
y_k = 
\underbrace{\begin{bmatrix} \mathsfbi{C} & 0 \end{bmatrix}}_{\tilde{\mathsfbi{C}}}
\begin{bmatrix}
\boldsymbol{x}_k \\
x_{\text{i},k}
\end{bmatrix} + v_k.
\end{eqnarray}
\par
The cost functions for feedback and the Kalman filter are adjusted accordingly to include integral action through the augmented matrices $\tilde{\mathsfbi{A}}$, $\tilde{\mathsfbi{B}}$, and $\tilde{\mathsfbi{C}}$. The state penalty is modified to include an additional cost to the integral term
\begin{equation}
    \mathsfbi{Q}_\text{LQR,i}=\begin{bmatrix}
        \mathsfbi{C^\mathrm{T}C} & \mathsfb{0}\\
        \mathsfb{0} & q_\text{i}
    \end{bmatrix},
\end{equation} where $q_\text{i}$ is the penalty term for integral action and is tuned manually.
\par
The combined LQG block diagram with integral action is shown in figure \ref{fig: model}. Deployment is achieved by coupling the controller to the simulation. The plant is now the solution to the Navier-Stokes equations, with output $y_k \,$=$\,C_\text{l}$ at each time step. The input $u_k$ is computed, and the velocity boundary condition at the actuator slots is updated according to equation \ref{eq:control} at each time step.
\section{Results}\label{sec: Results}
\subsection{System identification}
For model training, we use a snapshot matrix consisting of a mean-reduced flow field of the velocity magnitude $\boldsymbol{u'}\,$=$\,\boldsymbol{u}-\boldsymbol{\bar{u}}$, in the discretised domain $\upOmega$ where $\boldsymbol{u}\,$=$\,\sqrt{u_x^2+u_y^2}$ and
\begin{equation}
\left. \begin{array}{ll}
\displaystyle ~~~ -2\leq x\leq6\\[8pt]
\displaystyle  -1.5 \leq y\leq1.5 
 \end{array}\right\} \quad \in \quad \upOmega.
  \label{symbc}
\end{equation}
\par
Although other scalar fields, such as total pressure or spanwise vorticity, may be used for training, velocity magnitude was chosen as it has previously been identified as a particularly good variable for the decomposition of periodic cylinder flow \citep{Yildirim_2009}. To generate the training dataset, the URANS simulation is initialised with the flow field in the steady limit cycle of vortex shedding. A bandwidth-limited pseudo-random signal is applied to the control surfaces for broadband excitation and captures the forced response with a velocity signal amplitude standard deviation of $\upSigma_\text{u}\,$=$\,0.13U_\infty$. 
\par
\begin{figure}[h!]
\centerline{ 
\includegraphics[width=\textwidth]{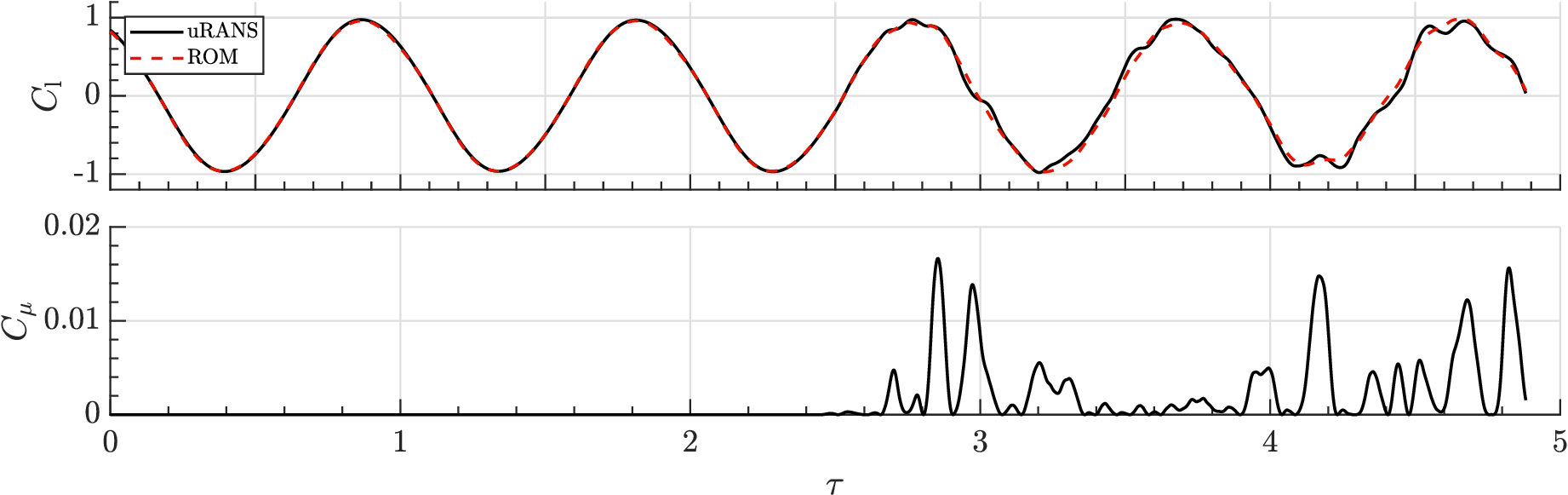}}
\caption{Evolution of lift coefficient compared to ROM prediction subject to bandwidth-limited actuation inputs.}\label{fig: lift}
\end{figure}

Since the actuation signal is always band-limited in practice, by virtue of finite actuation bandwidth, the pseudo-random signal is generated 
with a third-order Butterworth filter with a cut-off at $4f_\text{s}$ to identify a model that captures the fundamental vortex shedding mode and its subsequent 3 harmonics. The training data is 5 shedding cycles long. The time history of $C_\text{l}$ from the URANS simulation is compared to the ROM prediction for an $11{th}$-order model, given the state vector ($\boldsymbol{x}_k$) used for model training, in figure \ref{fig: lift}. The actuation inputs used for training are included. The actuation magnitude is normalised by the jet momentum coefficient ($C_\mu$) defined as
\begin{equation}
C_\mu=\frac{2\dot{m}_\text{c}{u_\text{c}}}{\rho U_\infty^2D}=\frac{2u_\text{c}^2s_\text{c}}{U_\infty^2 D},
\end{equation}
where $\dot{m}_\text{c}$ is the mass flow rate from the actuation slot, and $s_\text{c}$ is the area of the actuation slot. 
\subsection{Reduced-order modelling}
DMD \citep{Schmid_2010, Kutz_2016} provides a finite-dimensional estimate of a dynamic system's Koopman operator that models the behaviour of a non-linear system with an infinite-dimensional linear space. Therefore, by necessity, DMD finite-dimensional models are only an approximate linear representation of the original system. The truncation rank, $r$, determines the accuracy of the subsequent linear model, and as $r$ increases, the model captures more variance in the training data, resulting in improved system approximations.
\par
\begin{figure}[h!]
    \centering
    \includegraphics[width=\textwidth]{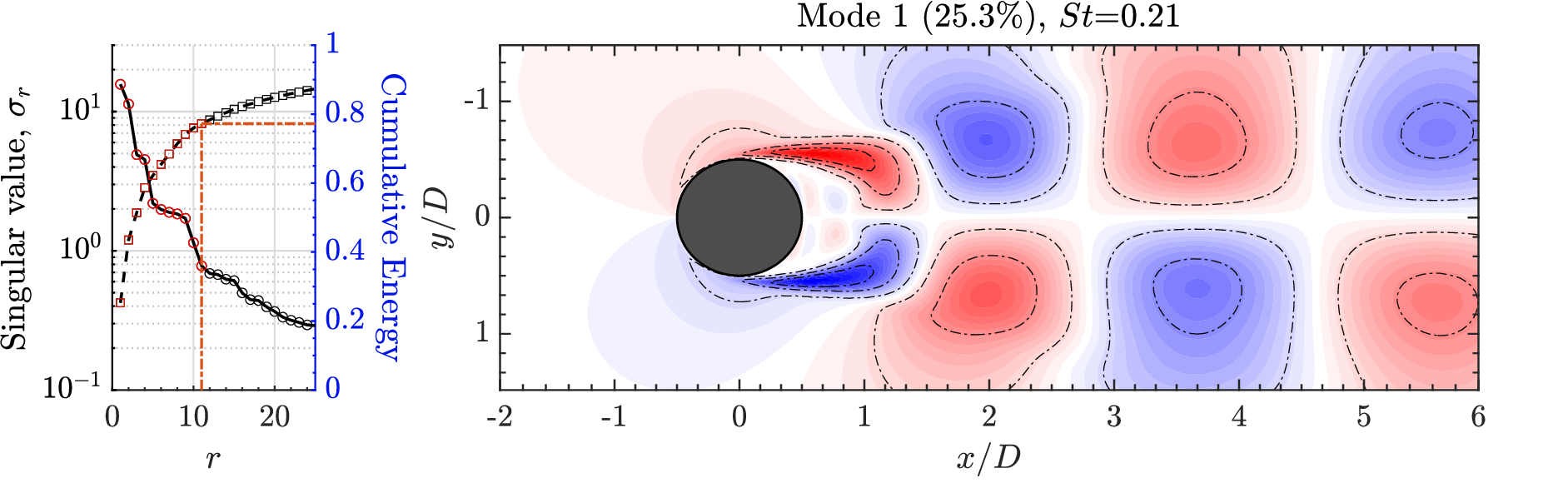}
    \caption{(a) Decay of singular values ($-\circ$) and cumulative energy ($--\square$), the first 11 modes are highlighted, capturing 77\% of the cumulative energy of the system. (b) Real part of the first DMD mode ($\upPhi_1$) with associated frequency matching the vortex shedding frequency, $St\,$=$\,0.21$.} \label{fig: spectrum}
\end{figure}

Figure \ref{fig: spectrum} illustrates the decay in singular values, $\sigma_r$, of each mode as the truncation rank increases and the cumulative normalised energy for each model size. The real part of the first spatial DMD mode ($\boldsymbol{\upPhi}_1$) is shown in figure \ref{fig: spectrum}. Each DMD mode is associated with an oscillatory frequency; the first mode's frequency is equal to the vortex shedding frequency, i.e. $St\,$=$\,0.21$. 
\par
Whilst one might be tempted to choose model order ($r$) to match, say, $99\%$ of energy, this is not necessarily a very good idea for any practical control design for several reasons. As can be seen from figure \ref{fig: spectrum}, the sharp decay in singular values suggests that as the model order increases, there is a decreasing energy contribution of the extra modes to the dynamics of the system. As higher-order modes are also at higher frequencies, any attempt to control them would place additional demands on the controller and, more importantly, on the realisable actuation bandwidth that fundamentally limits achievable performance. Conversely, including only very few modes in the model for control design risks failure as the unmodelled modes couple non-linearly into low-order modes, potentially triggering instability. 
\par
Fortunately, there exists a band gap in the frequency domain between vortex shedding and higher-order modes. This frequency separation between vortex shedding, harmonics and turbulent fluctuations is a well-known phenomenon in bluff-body flows, demonstrated experimentally \citep{Maryami_2020} and computationally \citep{elmiligui_2004}. We therefore hypothesise that URANS simulations are sufficient for creating suitably accurate ROM models with a small number of states for control design. URANS simulations damp out high-frequency turbulent fluctuations through turbulent eddy viscosity modelling whilst resolving the important vortex shedding dynamics. Since high-frequency turbulent fluctuation modes are not controllable with our postulated realistic actuation bandwidth, we treat them here as additional process noise, $w_k$, for feedback control design.
\subsection{Influence of model order on controller performance}
\begin{figure}[h!]
\centerline{ 
\includegraphics[width=\textwidth]{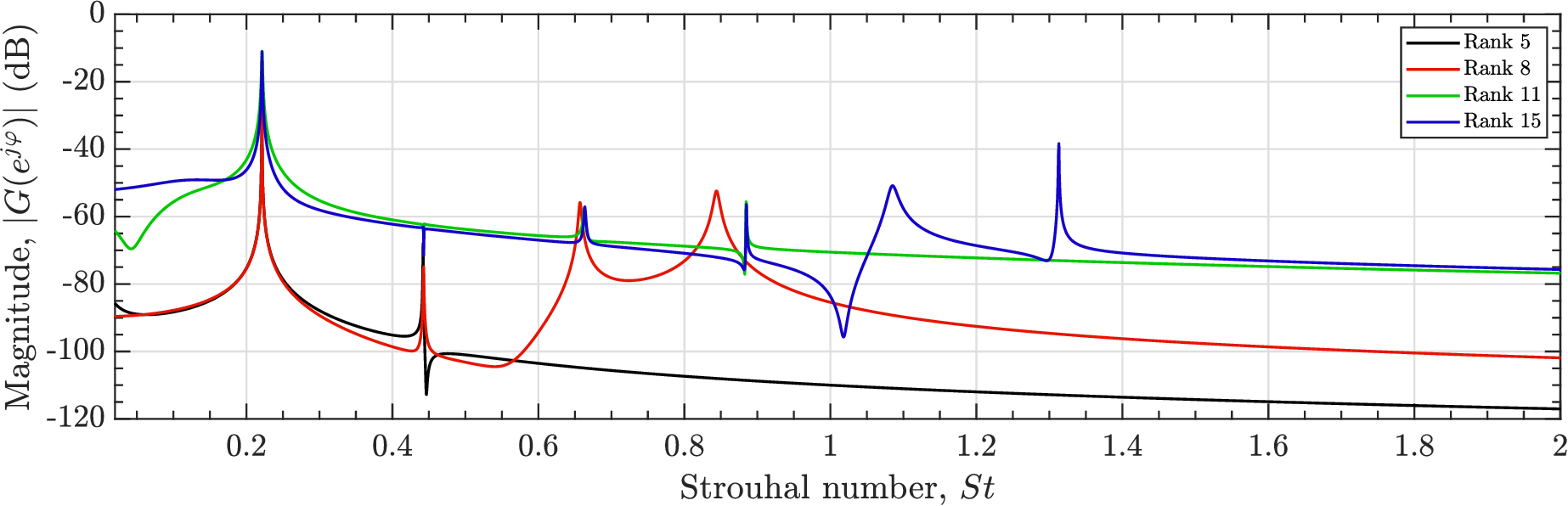}}
\caption{Frequency response of open-loop transfer functions for models of size: 5, 8, 11, and 15 against Strouhal number.}\label{fig: bode}
\end{figure}

Given state-transition ($\mathsfbi{A}$) and input-state ($\mathsfbi{B}$) matrices from DMDc and the output-state matrix ($\mathsfbi{C}$) described in \S\ref{sec: observe}, we may form a state-space model of size $r$. The open-loop transfer function magnitude of the frequency response for a range of model sizes is presented in figure \ref{fig: bode}.
\par
The dominant peak is fixed at the fundamental shedding frequency, $St\,$=$\,0.21$. As the truncation rank increases, higher-order vortex shedding modes are included. To compare the effect of model size on control performance, each feedback controller is tested with identical tuning parameters: $\mathsfbi{Q}_\text{LQR}\,$=$\,\mathsfbi{C^\mathrm{T}C}$ (thus output-weighted), $\mathsfbi{R}_\text{LQR}\,$=$\,3\times10^{-5}$. To quantify the effectiveness of a controller, a power-saving metric $\zeta$ is introduced, calculated as the difference between the power saved through drag reduction $\upDelta P_\text{d}$ and the power consumed through actuation $P_\text{act}$, normalised by the base drag power $P_\text{d,base}$.
\begin{equation}
    \zeta = \frac{\upDelta P_\text{d}-P_\text{c}}{P_\text{d,base}}\times100.
\end{equation}
The power saved by drag reduction is given as
\begin{equation}
    \upDelta P_\text{d}=P_\text{d,base}-P_\text{d,c}=(\langle F_\text{d,base} \rangle-\langle F_\text{d,c}\rangle)U_\infty,
\end{equation}
where $P_\text{d,base}$ and $P_\text{d,c}$ are the post-transient time-averaged drag power for the base and controlled flow, $\langle F_\text{d, base}\rangle$ and $\langle F_\text{d,c}\rangle$ are the time-averaged baseline and control drag over a sufficiently long time period, $\tau$\,$\in$\,$[20,40]$ for 2D simulations. The actuation power consumed is given as
\begin{equation}
P_\text{c}=\frac{1}{2}\rho \langle \lvert u_\text{c}\rvert^3 \rangle S_\text{c},
\end{equation}
where $\langle u_\text{c} \rangle$ is the time-averaged control jet velocity.
\par
This is not a true measure of a real system's power savings, as secondary power draws, such as electrical power for actuators and compressors, are not accounted for.
Importantly, feedback controllers have initial transient dynamics before reaching a lower drag solution (either fully stabilised or a new limit cycle). Therefore, controller performance metrics are taken in the time span after initial transients have settled. Whilst the controller observable is $C_\text{l}$, the ultimate goal of this approach is to reduce drag, thus motivating this performance metric.
\begin{figure}[h!]
\centerline{ 
\includegraphics[width=\textwidth]{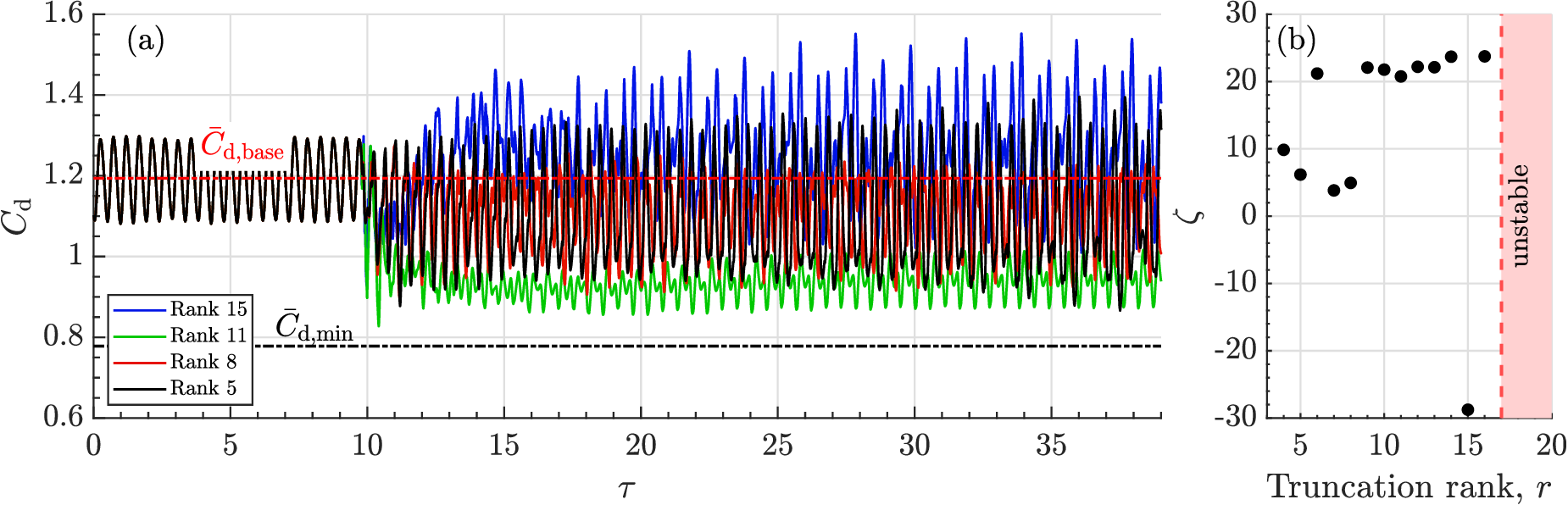}}
\caption{(a) Time history of drag coefficient for different model truncation ranks. Control is turned on at $\tau\,$=$\,10$. Average uncontrolled drag and minimum steady drag are included. (b) Performance, $\zeta$, of controllers with different truncation ranks.}\label{fig: performance}
\end{figure}
\par
The simplest possible model ($r\,$=$\,2$) captures the correct vortex shedding peak at $St\,$=$\,0.21$ using two complex conjugate modes. The rank 2 model contains no information about higher-order modes or the influence of the actuator on the flow field, and so the subsequent LQG controller design did not show a reduction in vortex shedding in closed-loop URANS simulations. The lowest-rank model that leads to successful controller design that suppresses lift oscillations and reduces drag is the $4{th}$-order model. However, the time history of the drag coefficient calculated through closed-loop simulations shows significant unsteadiness for models up to $r\,$=$\,8$ (see figure \ref{fig: performance}). Whilst $\zeta\,$=$\,10\%$ for the $4{th}$-order model, it is clear that higher-order models can improve on this result.
\par
\begin{figure}[h!]
\centerline{ 
\includegraphics[width=\textwidth]{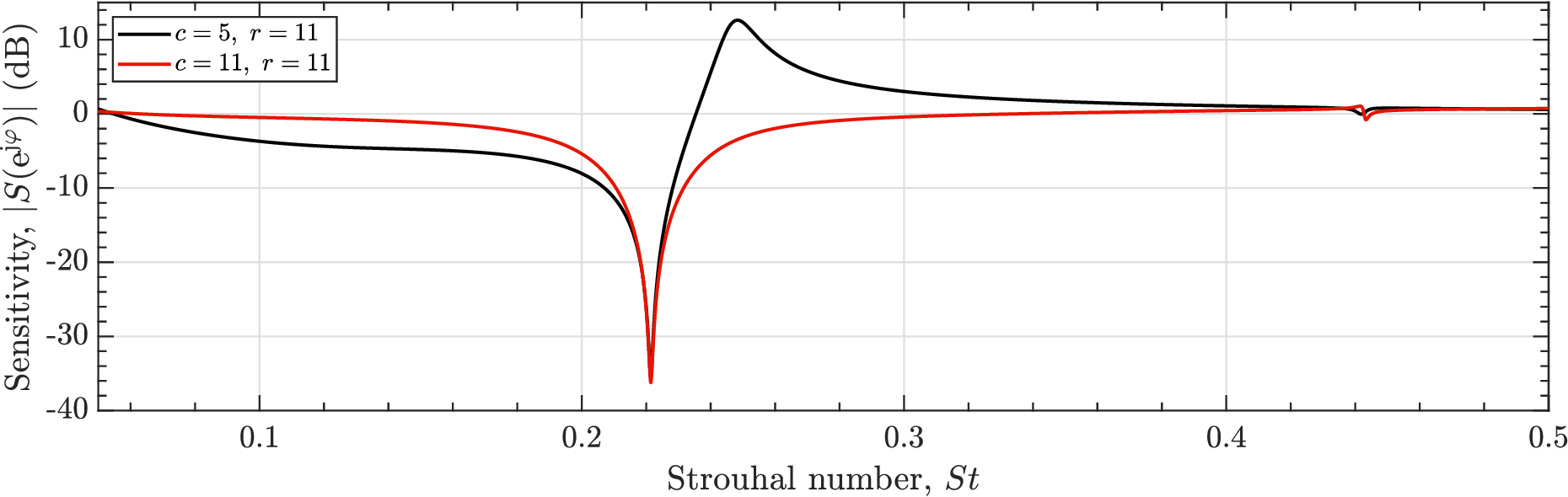}}
\caption{The sensitivity transfer functions for our system with an $11{th}$-order model but a controller of smaller ($5$) and equal size, highlighting instability due to unmodelled dynamics.}\label{fig: lowC}
\end{figure}

We propose that the reason for poor performance with smaller model sizes is due to the actuators exciting unmodelled higher-order modes, which destabilise the flow field and limit performance. This is demonstrated in figure \ref{fig: lowC}, showing the sensitivity transfer function for an $11{th}$-order plant and controller compared to a $11{th}$-order plant with a $5{th}$-order controller. The sensitivity transfer function for the system using a $ 5{th} $-order controller with a higher-order plant shows a reduction in gain around the vortex shedding frequency, but amplifies frequencies just outside this frequency band. This is the result of attempting to control a plant with unmodelled dynamics, which are unintentionally excited by a low-order controller.
\par
There appears to be an ideal range of truncation rank $9$\,$\leq$\,$r$\,$\leq$\,$14$ in which the power saved is $\zeta$\,$\approx$\,$20\%$. It is clear from figure \ref{fig: bode} that the improved performance of controllers in this size range is driven by the increased gain in frequency response at the vortex shedding peak compared to lower-order models.
\par
Increasing the model size further leads to negligible performance gains until the closed-loop performance starts to deteriorate and ultimately leads to instability. Although, strictly speaking, the closed-loop system is non-linear, the performance degradation of higher-order controllers can be explained through the use of the Bode integral theorem (\citet{Mohtadi_1990}) 
\begin{equation}
   \int^\pi_0\ln \lvert S(e^{j\varphi})\rvert d \varphi=\upi\sum^m_{i=1}\ln\lvert(\beta_i)\rvert,
   \label{eq: bode}
   \end{equation}
where $\beta_i$ are the unstable open-loop poles of the system, $m$ is the total number of these unstable poles, and $S(e^{j\varphi})$ is the sensitivity transfer function. The controller is designed to suppress instabilities by reducing the sensitivity gain $\lvert S(e^{j\varphi})\rvert$, in the vicinity of the eigenfrequencies of all the modelled modes represented by the DMD model.
\par
\begin{figure}[h!]
\centerline{ 
\includegraphics[width=\textwidth]{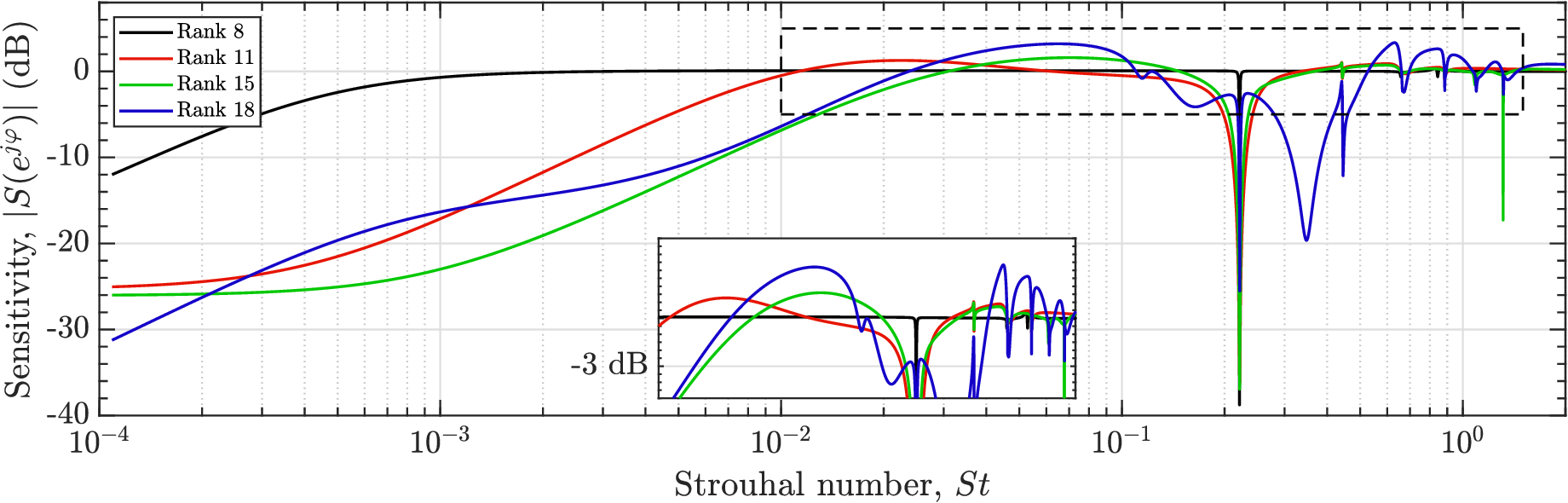}}
\caption{Sensitivity transfer function of models of rank $8,11,15$ and $18$ showing the waterbed effect increasing sensitivity to disturbances outside of $f_\text{s}$, resulting in increasingly unstable controllers.}
\label{fig: sensitivity}
\end{figure}

As a result of the Bode integral theorem, the attempted suppression of higher-frequency modes results in amplification of disturbances at other frequencies. This is the well-known waterbed effect, and can be seen in the sensitivity responses for models with increasing truncation rank in figure \ref{fig: sensitivity}. The rank 15 and 18 models include high-frequency peaks, which are not present in lower-order models. By attempting to suppress higher-frequency instabilities, the resulting waterbed effect pushes up sensitivity gains above $0$\,$\text{dB}$ at a range of other frequencies outside those resonant peaks, particularly in the broad frequency range $St$\,$\in$\,$[0.02,0.2]$ and $St$\,$>$\,$0.5$. If the sensitivity gain amplifies those disturbances, then any secondary instabilities that are present in those regions as a result of non-linear interactions with the modelled modes will get amplified, causing further non-linear coupling and ultimately leading to the instability of the overall closed-loop scheme. 
\par
To put it another way, large controller gains necessary to stabilise higher frequency modes of the linear system for which the controller is designed fundamentally create stronger non-linear interactions in the ``real" (e.g. closed-loop simulated system). This explains the significant drop in $\zeta$ for a rank $15$ model in figure \ref{fig: performance} and complete instability in models with truncation rank ($r$\,$>$\,$16$). The strong non-linearity of the problem therefore motivates the need for non-linear approaches that are currently being pursued \citep{Proudfoot_26}. For this present study, we have chosen a model with truncation rank $r\,$=$\,11$ that reflects a good balance of sufficient complexity to capture the main governing dynamics and $77\%$ of the system's energy, whilst leading to a well-behaved sensitivity function response.
\par
We may interpret which fluid structures are modelled through an eigenvalue analysis of the DMD modes which constitute the $11{th}$-order ROM. The first 2 DMD modes are complex conjugate pairs associated with vortex shedding. The subsequent 6 modes are similarly complex pairs, representing harmonics of the vortex shedding mode. Vortex shedding and subsequent harmonic modes are almost purely oscillatory, i.e. the eigenvalues sit on the unit circle. The remaining 3 modes represent interactions between the actuators and the wake. Modes 9 and 10 are under-damped oscillatory pairs, and mode 11 is over-damped (real eigenvalue). A summary of the eigenvalues associated with each mode in the $11{th}$-order model is expressed as damping ratios and Strouhal numbers in table \ref{tab:model_poles}.
\begin{table}
  \begin{center}
\def~{\hphantom{0}}
  \begin{tabular}{lccc}
      Mode No.  & Pole   &   Damping ratio & $\mathrm{St}$ \\[3pt]
       1,2   & $9.99$\,$\times$\,$10^{-1}$\,$\pm$\,$8.13$\,$\times$\,$10^{-3}i$  & ~~$2.24$\,$\times$\,$10^{-3}$~ & 0.21 \\
       3,4   & $9.99$\,$\times$\,$10^{-1}$\,$\pm$\,$1.63$\,$\times$\,$10^{-2}i$    & ~~$4.46$\,$\times$\,$10^{-4}$~ & 0.42 \\
       5,6   & $9.99$\,$\times$\,$10^{-1}$\,$\pm$\,$2.43$\,$\times$\,$10^{-2}i$    & ~~$2.12$\,$\times$\,$10^{-3}$~ & 0.63 \\
       7,8   & $9.99$\,$\times$\,$10^{-1}$\,$\pm$\,$3.25$\,$\times$\,$10^{-2}i$    & ~~$2.94$\,$\times$\,$10^{-4}$~ & 0.84 \\
       9,10  & $9.97$\,$\times$\,$10^{-1}$\,$\pm$\,$4.13$\,$\times$\,$10^{-3}i$  & ~~$5.33$\,$\times$\,$10^{-1}$~ & 0.13 \\
       11    & $9.98$\,$\times$\,$10^{-1}$  & ~~$1.00$  ~ & 0
  \end{tabular}
  \caption{Pole locations for discrete $11{th}$-order model with associated damping and Strouhal number.}   \label{tab:model_poles}
  \end{center}
\end{table}
\par
To verify the prediction accuracy of the $11{th}$-order model, we investigate the ROM-predicted response compared to simulation results. The actuation is driven open-loop with phase-shifted sinusoidal forcing according to
\begin{equation}
u_\text{c}(t)=0.6U_\infty \sin(2\upi f_\text{s}t+9\upi/10).
\end{equation}
\par
\begin{figure}[h!]
\centerline{ 
\includegraphics[width=\textwidth]{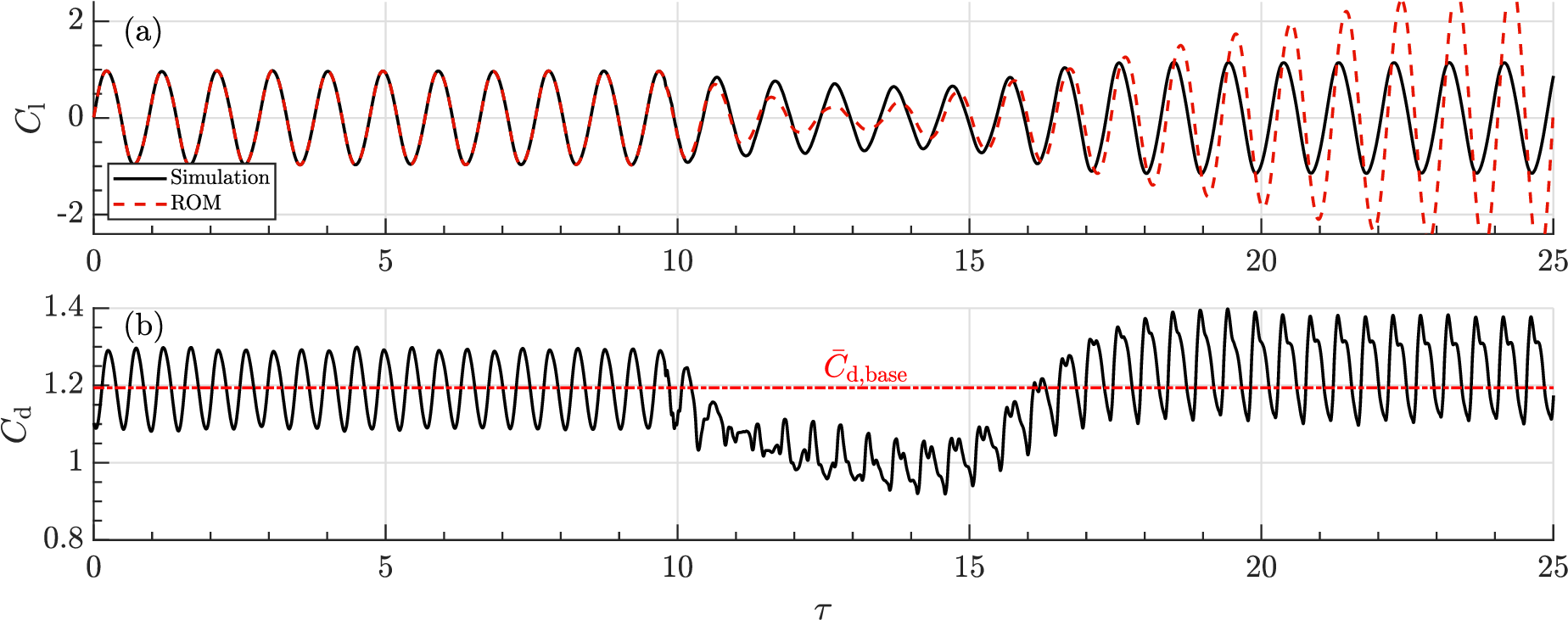}}
\caption{(a) URANS simulation and $11{th}$-order ROM prediction of $C_\text{l}$ subject to open-loop phase-shifted actuation turning on at $\tau\,$=$\,10$. $u_\text{c}(t)\,$=$\,0.6U_\infty\sin(2\upi f_\text{s}t+9\upi/10)$, the reduced-order model predicts large unsteady growth due to resonant forcing. (b) $C_\text{d}$ from URANS increases with open-loop forcing due to the coupling of actuation with vortex shedding.}\label{fig: sin}
\end{figure}

The response of the lift coefficient to this forcing function is illustrated in figure \ref{fig: sin}, comparing open-loop predictions against simulation. The drag coefficient from URANS is included, demonstrating the coupling between reducing lift variance and minimising drag.
\par
Sinusoidal forcing illustrates why open-loop schemes are unlikely to succeed. If the system is forced in open-loop at $\mathrm{St}\,$=$\,0.21$ with the correct phase difference to oppose vortex shedding, the control scheme is initially successful at suppressing shedding. However, as suppression occurs, the frequency of shedding changes, resulting in a phase shift between input and shedding. The phase difference increases until the vortex shedding locks in-phase with the actuation signal, resulting in an unstable increase in overall oscillations and a subsequent rise in drag. The linear model predicts that this would continue indefinitely (undamped resonance), when in reality, non-linear coupling in the flow results in a new, higher drag limit cycle. Whilst the ROM accurately predicts this qualitative behaviour, it can be seen that the magnitude of response to actuation is overestimated.
\subsection{Control Performance}\label{sec :performance}
\par
To evaluate the controller's performance compared to an optimal (fully stabilised) solution, we introduce an idealised minimum obtainable drag from a symmetric half-domain simulation $\bar{C}_\text{d,min}$. This setup prevents any instability from initiating, providing a lower bound to aerodynamic drag, and thus an optimal control policy, as our scheme may only reduce unsteady pressure drag. The maximum possible drag reduction therefore corresponds to a complete reduction of the mean flow correction, as proposed in \citet{protas_2002}. The minimum drag was found to be $\bar{C}_\text{d,min}\,$=$\,0.78$.
To measure controller performance, a normalised average post-transient drag reduction is used ($\bar{C}_\text{d,c}$), where post-transience is taken to be $\tau$\,$\geq$\,$20$. $\eta$ is identical to that used by \citet{xia_2024} and represents the fraction between achieved drag reduction and maximum possible drag reduction,
\begin{equation}
    \eta = \frac{\bar{C}_\text{d,base}-\bar{C}_\text{d,c}}{\bar{C}_\text{d,base} -\bar{C}_\text{d,min}}\times100\%.
\end{equation}
\par
\begin{figure}[h!]
\centerline{ 
\includegraphics[width=\textwidth]{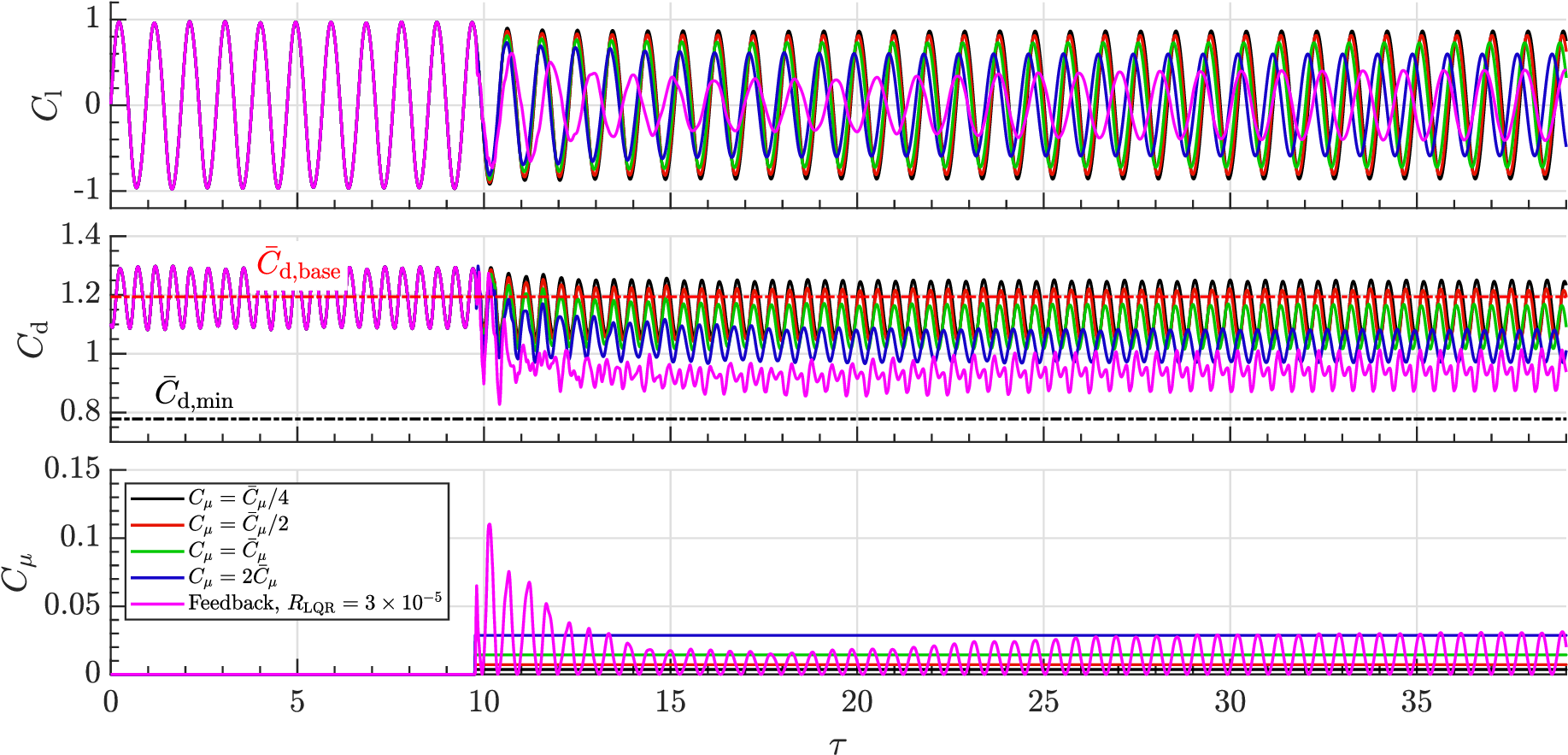}}
\caption{Time evolution of lift and drag coefficients comparing the LQG controller to constant blowing. Actuation provides the same $\bar{C}_\mu$ for each case.}\label{fig: constant}
\end{figure}

To quantify the importance of feedback, we compare the performance of the controller against constant DC blowing in the range $\bar{C}_{\mu,11}/4$\,$\leq$\,$C_{\mu,\text{DC}}$\,$\leq$\,$2\bar{C}_{\mu,11}$, where $\bar{C}_{\mu,11}$ is the post-transient average momentum coefficient for the feedback case with a $11{th}$-order controller and $C_{\mu,\text{DC}} $ is the momentum coefficient for the constant blowing case. A summary of constant forcing results is presented in table \ref{tab: DC}.
\begin{table}
  \begin{center}
\def~{\hphantom{0}}
  \begin{tabular}{lccc}
      Case  & $\bar{C}_\mu$ & $\zeta$ & $\eta$ \\[3pt]
       Feedback, ($r$\,$=$\,$11$) & 0.0141  & 20.8\% & 60.6\% \\
       DC, ($\bar{C}_{\mu,11}/4$) & 0.0035  & 3.0\%& 8.7\%\\
       DC, ($\bar{C}_{\mu,11}/2$) & 0.0071  & 4.9\%& 14.4\%\\
       DC, ($\bar{C}_{\mu,11}$) & 0.0141  & 8.2\% & 24.4\%\\   
       DC, ($2\bar{C}_{\mu,11}$) & 0.0282  & 13.1\%& 40.2\%\\
  \end{tabular}
  \caption{Summary of performance metrics comparing feedback $11{th}$-order controller against DC blowing for a range of momentum coefficients.}   \label{tab: DC}
  \end{center}
\end{table}
For feedback, the controller is tuned with the parameters; $\mathsfbi{Q}_\text{LQR}$\,$=$\,$\mathsfbi{C}^\mathrm{T}\mathsfbi{C}$, $\mathsfbi{R}_\text{LQR}$\,$=$\,$3\times 10^{-5}$. Direct measurements of states would require continuous knowledge of the velocity magnitude field, which is not viable for practical applications as discussed in \S \ref{sec: observe}. A Kalman filter is therefore used for state estimation with tuning parameters $\mathsfbi{Q}_\text{f}$\,$=$\,$1$, $\mathsfbi{R}_\text{f}$\,$=$\,$3\times 10^{-4}$. Tuning the controller requires several iterations to find the correct balance of actuator power and system stability, but may be guided by ROM predictions. The time evolution of $C_\text{l}$, $C_\text{d}$, and $C_\mu$ is illustrated in figure \ref{fig: constant} comparing the $11{th}$-order feedback controller to constant blowing cases.
\par
With feedback, a value of $\eta$\,$=$\,$60.6\%$ is achieved and $\zeta $\,$=$\,$ 20.8\%$. If we consider constant blowing using an equal level of momentum coefficient, we can only achieve performance at $\eta$\,$=$\,$24.4\%$ and $\zeta$\,$=$\,$8.2\%$. We may conclude that the introduction of feedback provides a significant benefit in reducing vortex shedding instabilities, with a greater overall reduction and for a lower actuation cost. In fact, the majority of the feedback controller's cumulative energy expenditure on actuation is spent during initial transience. Once a smaller limit cycle is reached, it is significantly cheaper to maintain stability. Reducing the magnitude of constant blowing does not improve efficiency. Indeed, at lower magnitudes of blowing, power efficiency is lower, with worse suppression of vortex shedding. Increasing the degree of blowing reduces vortex shedding, but power consumption is far greater, confirming that a lack of feedback will result in a less efficient solution.
\subsection{Influence of controller tuning}
\label{sec: R sweep}
\begin{figure}[h!]
\centerline{ 
\includegraphics[width=\textwidth]{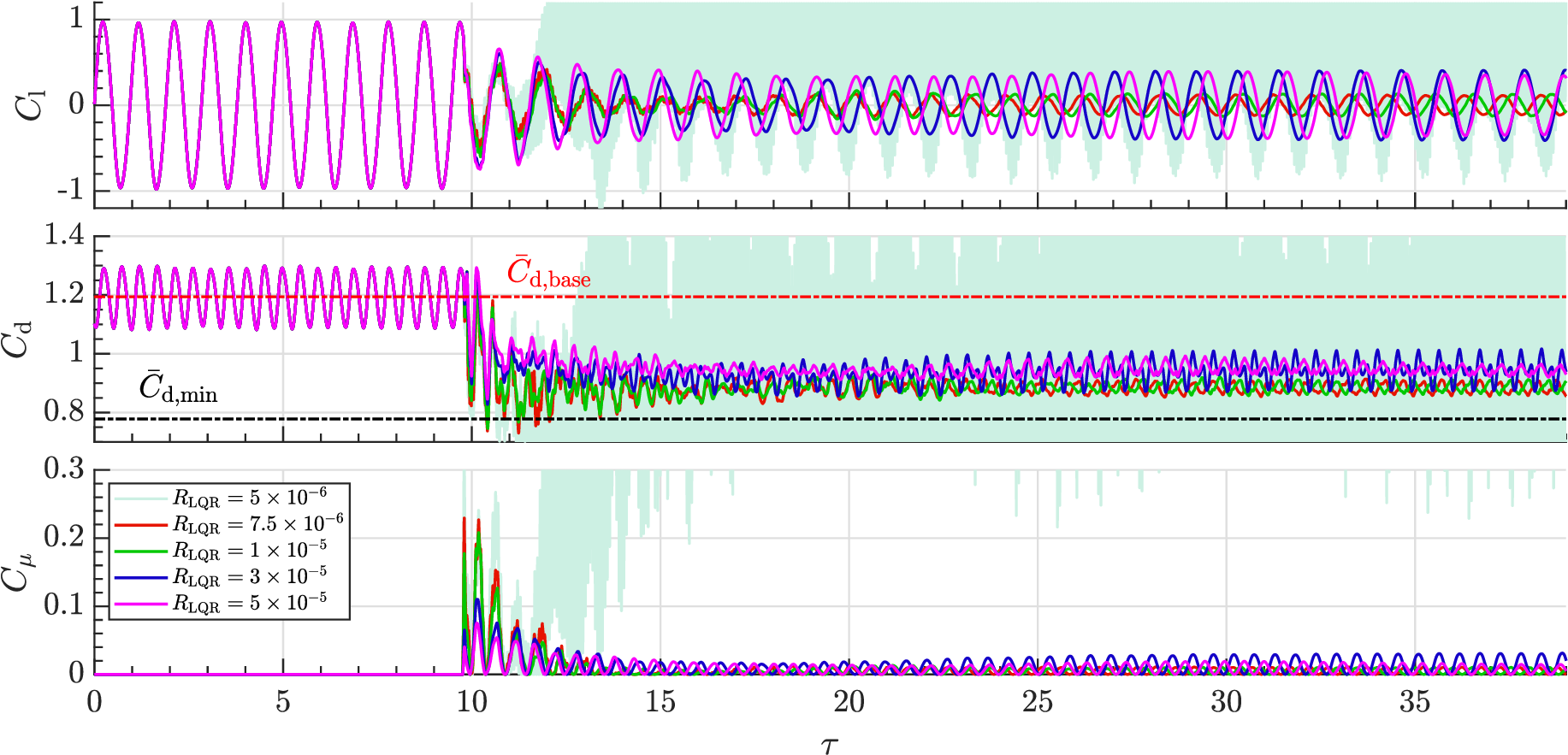}}
\caption{Time evolution of lift and drag coefficients for controllers of the same rank but with changing actuator penalisation, $\mathsfbi{R}_\text{LQR}$.}\label{fig: rsweep}
\end{figure}

The linear quadratic regulator used for feedback is optimal with respect to minimising the cost function in equation \ref{eq: costfnc}. Controller performance may improve with a reduction in the actuation penalty ($\mathsfbi{R}_\text{LQR}$). This reduces the relative cost of actuation compared to state deviation, resulting in a more aggressive controller with larger actuation magnitudes. A range of controllers based on an $11{th}$-order model were tested varying $\mathsfbi{R}_\text{LQR}$ in the range
$5$\,$\times$\,$10^{-5}$\,$\leq$\,$\mathsfbi{R}_\text{LQR}$\,$\leq$\,$5$\,$\times$\,$10^{-6}$, holding all other controller tuning parameters constant. The results of these tests are presented in figure \ref{fig: rsweep}. 
\par
By reducing the relative penalisation of actuation, the controller's absolute $C_\mu$ increases, resulting in greater suppression of the shedding instabilities. However, actuation may not increase indefinitely; physical actuators are subject to rate and saturation limits. Furthermore, the control model is obtained by a linearisation about the shedding limit cycle. The true cylinder instability contains a non-linear transition from the unstable fixed point to the limit cycle with a cubic damping term \citep{Provansal_Mathis_Boyer_1987}. As actuation penalty reduces, actuation velocities increase for non-linear fluid interactions to become significant.  Consequently the model's linearity assumptions break down, resulting in worse overall performance. 
\par
\begin{figure}
  \centering
  \begin{minipage}[t]{0.47\textwidth}
    \vspace{0pt}
    \centering
    \def~{\hphantom{0}}

    \resizebox{\linewidth}{!}{%
    \begin{tabular}{lccccc}
      \hline
      $\mathsfbi{R}_\text{LQR}$ & $\bar{C}_\mu$ & $\hat{C}_\mu$ & $\bar{C}_\text{d}$ & $\zeta$ & $\eta$ \\[3pt]
      \hline
      $5\times10^{-5}$   & $0.0082$ & $0.075$ & $0.95$ & $20.4\%$   & $59\%$ \\
      $3\times10^{-5}$   & $0.0141$  & $0.11$  & $0.94$ & $20.8\%$   & $60.6\%$ \\
      $2\times10^{-5}$   & $0.0100$ & $0.15$  & $0.9$  & $23.7\%$   & $68.8\%$ \\
      $1\times10^{-5}$   & $0.0049$ & $0.21$  & $0.89$ & $25.6\%$   & $73.9\%$ \\
      $7.5\times10^{-6}$ & $0.0052$ & $0.23$  & $0.88$ & $25.9\%$   & $74.7\%$ \\
      $5\times10^{-6}$   & $1.79$   & $6.12$  & $0.74$ & $-355.6\%$ & $108.6\%$ \\
      \hline
    \end{tabular}%
    }
  \end{minipage}%
  \hfill%
  \begin{minipage}[t]{0.47\textwidth}
    \vspace{0pt}
    \centering
    \includegraphics[height=0.205\textheight,width=\linewidth,keepaspectratio]{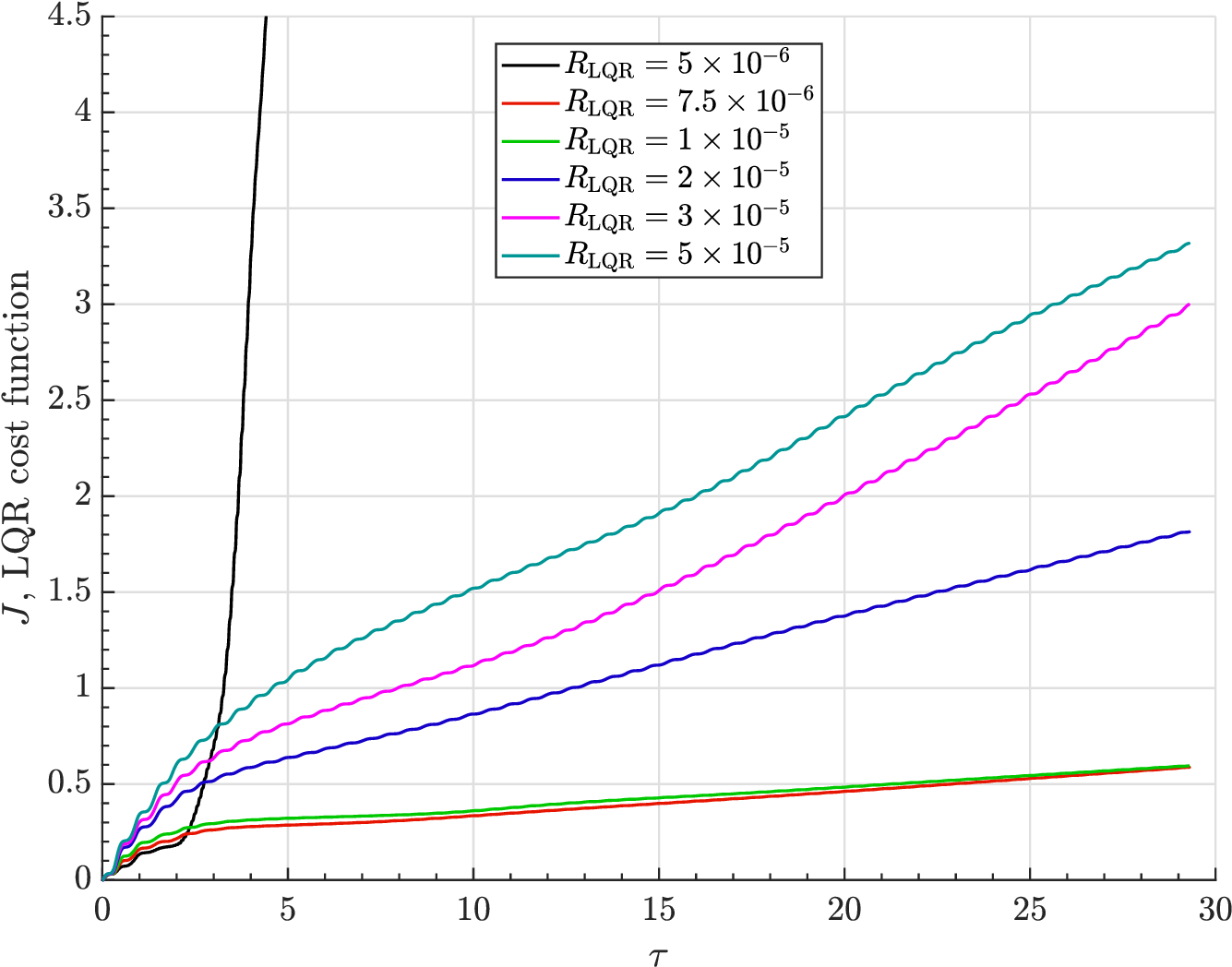}
  \end{minipage}

  \vspace{6pt}

  \begin{minipage}[t]{0.47\textwidth}
    \vspace{0pt}
    \captionsetup{type=table,width=\linewidth}
    \caption{Summary of the effect of reducing actuator penalty, $\mathsfbi{R}_\text{LQR}$.}
    \label{tab: R_sweep}
  \end{minipage}%
  \hfill%
  \begin{minipage}[t]{0.47\textwidth}
    \vspace{0pt}
    \captionsetup{type=figure,width=\linewidth}
    \caption{LQR cost function over non-dimensional time $J$\,$=$\,$ (\boldsymbol{x}_k^\mathrm{T}\mathsfbi{Q}_\text{LQR}\boldsymbol{x}_k$\,$+$\,$\mathsfbi{R}_\text{LQR}u_k^2)$ with varying actuator penalty.}    \label{fig: R_sweep}
  \end{minipage}
\end{figure}

A summary of the effects of reducing actuator penalty is provided in table \ref{tab: R_sweep}. As $\mathsfbi{R}_\text{LQR}$ reduces, $\zeta$ and $\eta$ increase. The best-performing controller achieves a power-saving performance of $\zeta$\,$=$\,$25.9\%$ and reaches $\eta$\,$=$\,$74.7\%$. Reducing $\mathsfbi{R}_\text{LQR}$ further results in actuation velocities so large that the linearity assumptions break down resulting in an unstable closed-loop. $C_\text{l}$ variance therefore greatly increases. Furthermore, as $\mathsfbi{R}_\text{LQR}$ reduces, the corresponding troughs in the closed-loop sensitivity at the model's designed frequencies will inevitably lead to increased sensitivity at other frequencies due to the waterbed effect (see equation \ref{eq: bode}). This effect is similarly destabilising.
\par
\begin{figure}[h!]
\centerline{ 
\includegraphics[width=\textwidth]{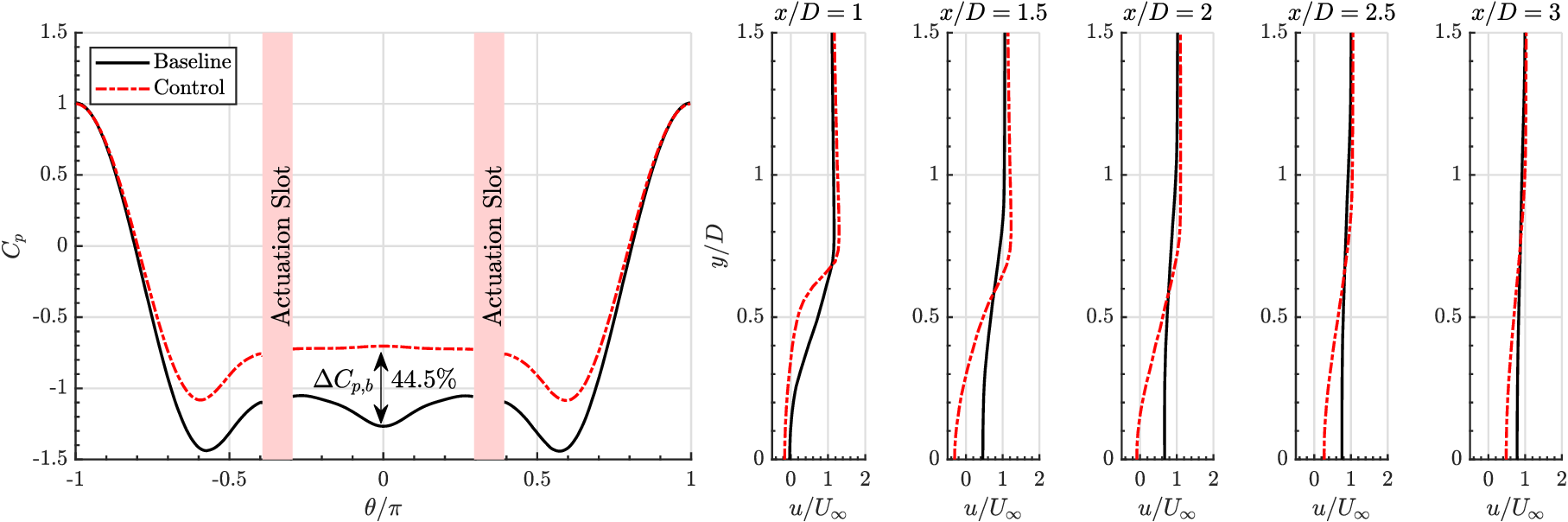}}
\caption{Influence of controller on pressure coefficient, $C_p$, across the cylinder surface and streamwise velocity wake profiles at several downstream locations from the trailing edge.}\label{fig: cp_theta}
\end{figure}

The effect of the best-performing controller, with $\mathsfbi{R}_\text{LQR}$\,$=$\,$7.5$\,$\times$\,$10^{-6}$, on the time-averaged pressure coefficient and streamwise velocity in the wake is presented in figure \ref{fig: cp_theta}. The variation in $C_p$ against cylinder angle ($\theta$), where $\theta/\upi$\,$=$\,$0$ is the trailing edge, shows that the feedback controller can improve the base-pressure coefficient by $44.5\%$. The controller can reduce the pressure loss throughout the cylinder's wake despite actuation slots being present in a small post-separation region, indicating that the control mechanism is indeed disrupting the formation of the vortex street. Average streamwise velocity, $u$, in the wake shows an elongated wake when feedback is active, attributed to the controller's stabilising influence, which corresponds to a lower drag limit cycle.
\par
The average post-transient momentum coefficient, $\bar{C}_\mu$, does not always increase as $\mathsfbi{R}_\text{LQR}$ reduces. This is because the initial transient control response increases, i.e. the maximum momentum coefficient, $\hat{C}_\mu$, increases as $\mathsfbi{R}_\text{LQR}$ reduces. This results in a rapid stabilisation of the wake, which subsequently requires less actuation to maintain, resulting in improved efficiency. Controller design is therefore a balance between efficiency and stability. Indeed, in figure \ref{fig: R_sweep} it is shown that the controller with the smallest cumulative cost is not the most aggressive or least aggressive controller. Instead, it is one with a strong initial transient response which rapidly stabilises the vortex shedding, then requires very little energy to maintain stability, highlighting the importance of respecting unmodelled non-linear instabilities.
\par
\begin{figure}[h!]
  \centering{
\includegraphics[width=\textwidth]{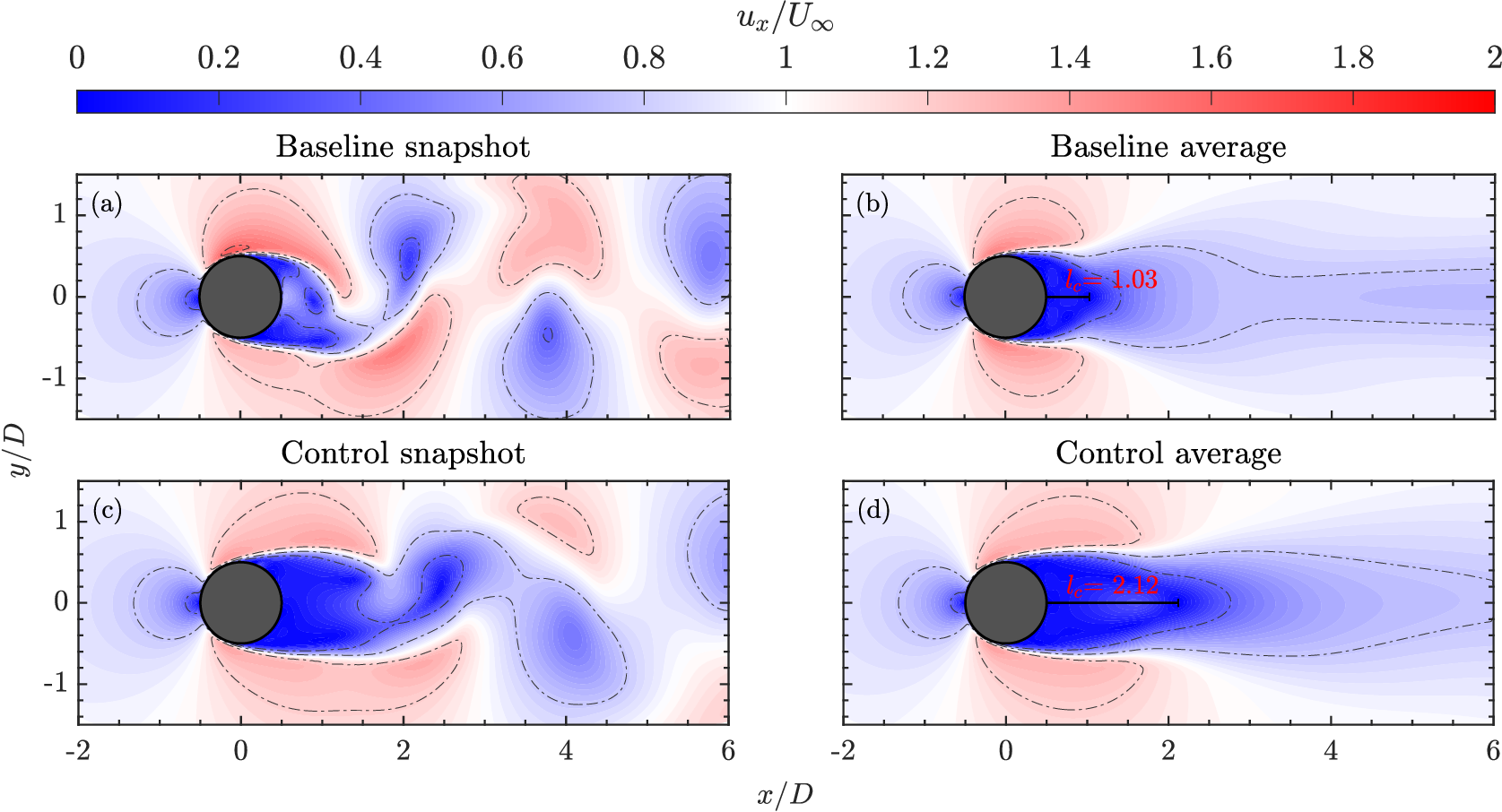}}
\caption{Contours of normalised streamwise velocity $u_x/U_\infty$ in the domain $\upOmega$\,$\in$\,$[-1.5$\,$\leq$\,$y$\,$\leq$\,$1.5],\newline [-2$\,$\leq$\,$x$\,$\leq$\,$6]$. Snapshots are at $\tau$\,$=$\,$40$, averaged: (a) Baseline snapshot, (b) Baseline average, (c) Control snapshot, (d) Control average. Control parameters: $\mathsfbi{Q}_\text{LQR}$\,$=$\,$\mathsfbi{C}^\text{T}\mathsfbi{C}$, $\mathsfbi{R}_\text{LQR}$\,$=$\,$7.5\times10^{-6}$, $\mathsfbi{Q}_\text{f}$\,$=$\,$1$, $\mathsfbi{R}_\text{f}$\,$=$\,$3\times10^{-4}$.}\label{fig: velocity}
\end{figure}

The reduction in vortex shedding can be observed visually by comparing the baseline flow field to the controlled flow field. Figure \ref{fig: velocity} presents a snapshot and mean contour of streamwise velocity for both baseline and controlled flow. The normalised recirculation bubble length, $l_\text{c}$, defined as the length from the cylinder's trailing edge to the point of zero average streamwise velocity on the centreline, is computed for baseline and controlled flow, with control doubling the recirculation length from $l_\text{c,base}$\,$=$\,$1.03$ to $l_\text{c,control}$\,$=$\,$2.12$. A longer recirculation length directly corresponds to lower average drag.
\subsection{Reference tracking and disturbance rejection}
As discussed in \S \ref{sec: ref}, the ability of the controller to track a desired non-zero reference signal is enabled through the implementation of integral action. To test the efficacy of integral action, the reference $C_\text{l,ref}$\,$=$\,$r(\tau)$ is changed according to the function
\begin{equation}
  r(\tau) = \left\{
    \begin{array}{ll}
      0, & 0 \le \tau\le 10 \\[2pt]
      0.09,   & 10 \le \tau\le 50 \\[2pt]
      -0.26,   &  \tau \ge 50.
    \end{array} \right.
\end{equation}
\par
\begin{figure}[h!]
\centerline{ 
\includegraphics[width=\textwidth]{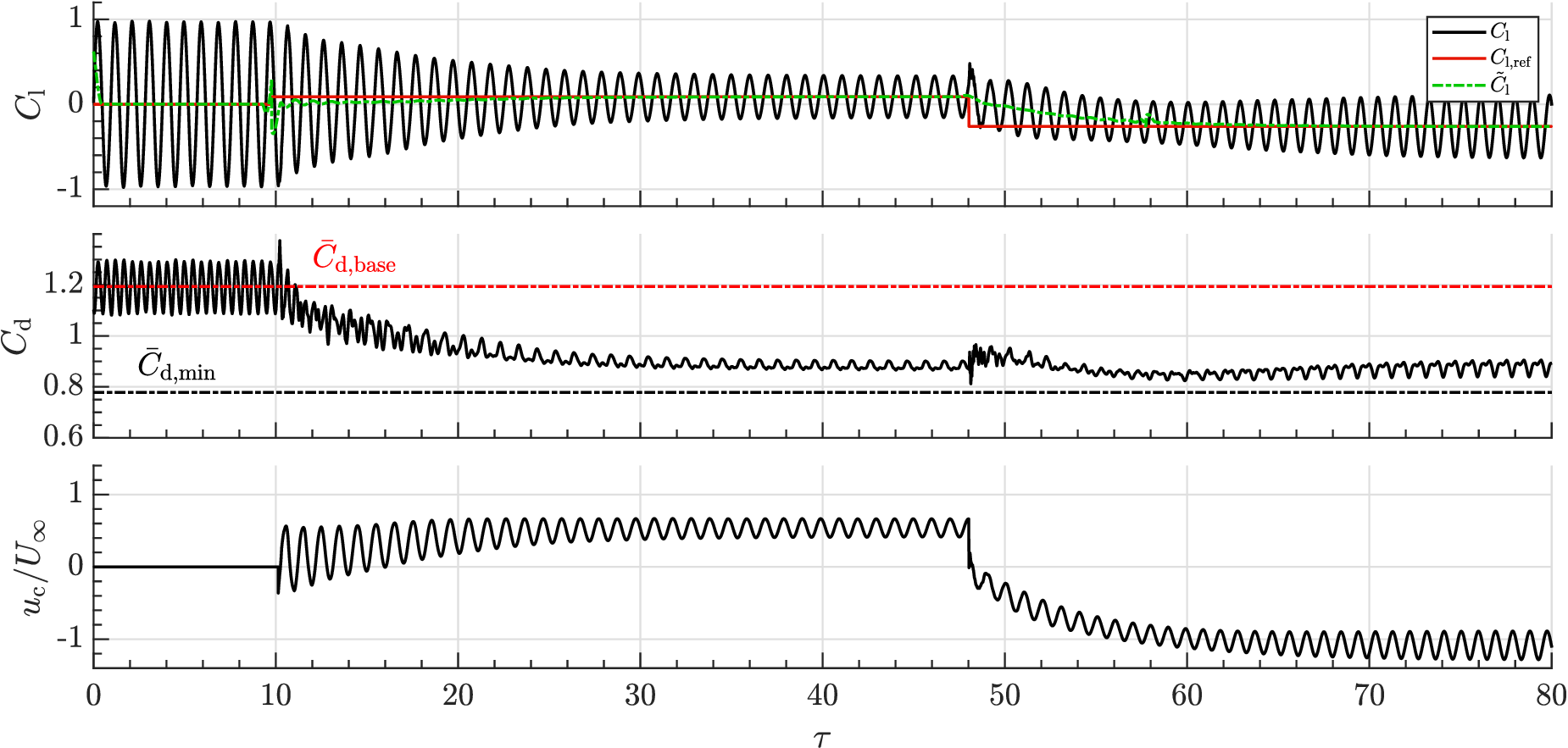}}
\caption{Time evolution of lift coefficient $C_\text{l}$ and normalised actuator velocities whilst tracking a desired reference; a moving average $\tilde{C}_\text{l}$ is provided. The controller is turned on at $\tau$\,$=$\,$10$. Drag coefficient and normalised inputs are also shown.}\label{fig: ref}
\end{figure}

The integrator gain is tuned manually to provide a satisfactory rise time for step inputs, leading to $q_\text{i}$\,$=$\,$5\times10^{-6}$. The resulting lift coefficient history is shown in figure \ref{fig: ref} for an $11{th}$-order controller. A moving average $\tilde{C}_\text{l}$ (window size\,$=$\,$1/f_\text{s}$) is provided. The controller successfully reduces lift variance whilst tracking the desired reference. The introduction of reference tracking comes with a cost of the need for greater actuation authority, as an increasing steady-state reference velocity results in larger average actuation velocities.
\par
To estimate the effect that real sensor noise would have on our design, synthetic Gaussian white noise is added to create a ``measured'' lift signal $\tilde{y}$\,$=$\,$y+\nu$ in the simulation. The sensor noise results in a signal-to-noise ratio of $9.6$\,dB when compared to the baseline lift signal. An additional low-frequency external disturbance is generated by allowing the inlet velocity to oscillate sinusoidally according to the function
\begin{equation}
    U_\infty(t)=U_\infty+0.1U_\infty \sin(2\upi f_\text{s}t/5).
\end{equation}
\par
\begin{figure}[h!]
\centerline{ 
\includegraphics[width=\textwidth]{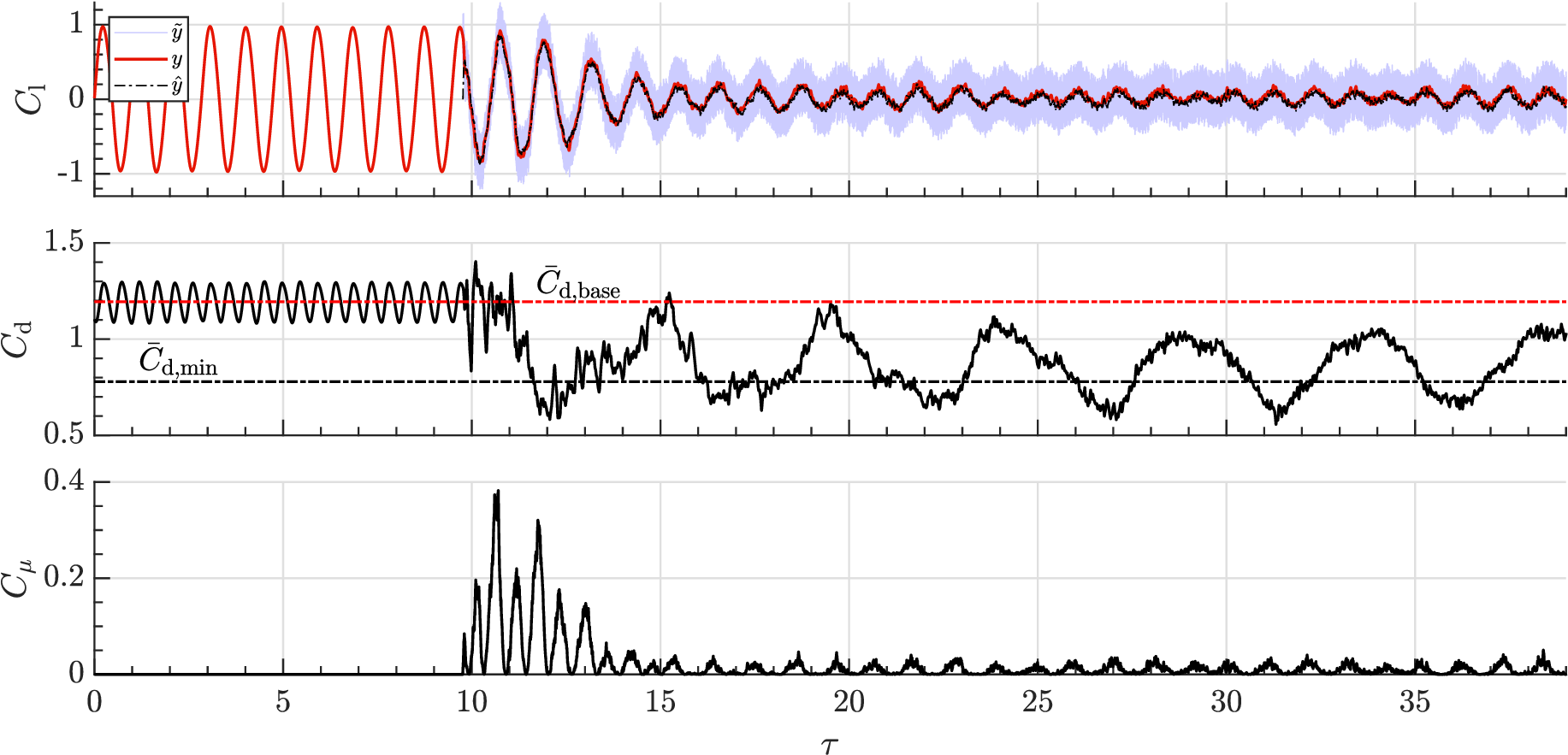}}
\caption{Time evolution of lift, drag, and momentum coefficient for the controlled plant subject to external disturbance and white sensor noise. The noisy lift, $\tilde{y}$, true lift, $y$, and the Kalman filter estimated lift, $\hat{y}$, are included.}\label{fig: noise}
\end{figure}

The evolution of the lift and drag coefficients over time is shown in figure \ref{fig: noise}. The noisy ($\tilde{y}$), true ($y$), and estimated ($\hat{y}$) lift coefficients are shown, with the Kalman filter accurately estimating lift through the noise and the controller successfully reducing lift variance. Drag reduction is accordingly achieved. Note that noise is filtered sufficiently without being passed through to the actuator above a given bandwidth, whilst large disturbances to the lift coefficient are rejected. The drag coefficient oscillates due to large streamwise global pressure variations in the simulation caused by the changing inlet boundary condition.
\par
\subsection{Robustness to Reynolds number}
\begin{figure}[h!]
\centerline{ 
\includegraphics[width=\textwidth]{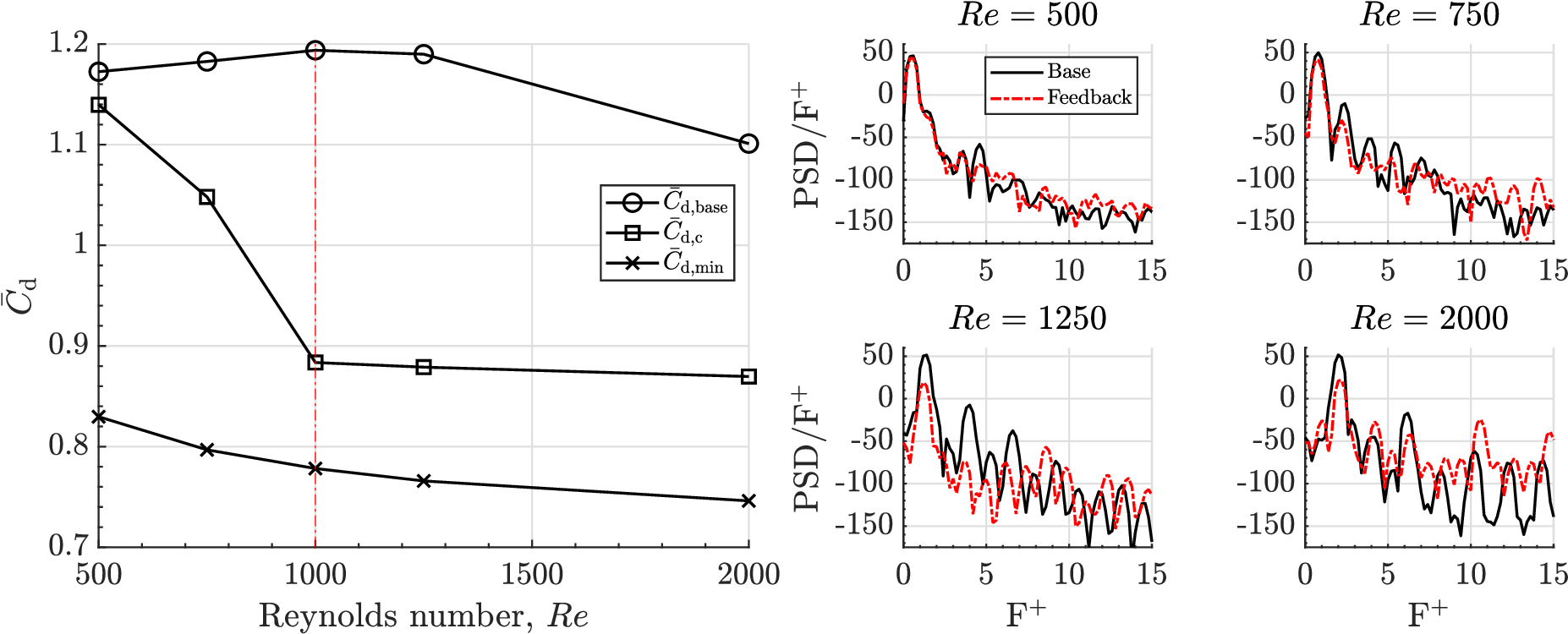}}
\caption{Post-transient drag coefficient $\bar{C}_\text{d}$ for baseline, minimum and controller flow over a range of Reynolds numbers. The controller was trained at $\Rey$\,$=$\,$1000$ (dashed line) and tested at $\Rey$\,$=$\,$500,~750,~1250,~2000$. Power spectral density of lift coefficient $C_\text{l}$ is presented for each off-design test, comparing baseline to controlled solutions over normalised frequency F$^+$.}\label{fig: reynolds}
\end{figure}

The controller, trained at $\Rey$\,$=$\,$1000$, is tested at different Reynolds numbers to study robustness to uncertainty in the operating conditions. Neither the simulation or controller time steps are changed. Figure \ref{fig: reynolds} shows the performance of the $11{th}$-order controller with $\mathsfbi{R}_\text{LQR}$\,$=$\,$7.5\times10^{-6}$ applied to these operating conditions, different to those used in training. The controller is still capable of reducing drag, achieving a performance of $\eta$\,$=$\,$9.6\%$, $34.9\%$, $73.4\%$, $65.2\%$ at $\Rey$\,$=$\,$500$, $750$, $1250$, $2000$, respectively. This compares to the performance of the best controller of $\eta$\,$=$\,$74.7\%$ at $\Rey$\,$=$\,$1000$ for which the controller was trained. Controller performance appears to degrade significantly at reduced Reynolds numbers, whereas controller performance is strong at higher speeds. The power spectra of lift coefficient $C_\text{l}$ are presented against normalised frequency, $\text{F}^+$\,$=$\,$f/f_\text{s,b}$, where $f_\text{s,b}$ is the shedding frequency at $\Rey$\,$=$\,$1000$. Whilst the controller does not appear to be capable of reducing the vortex shedding peak at $\Rey$\,$=$\,$500$ and $750$, it is relatively insensitive to higher Reynolds numbers. However, at $\Rey$\,$=$\,$2000$, the controller clearly begins to excite higher-frequency instabilities, which would explain the reduced performance. These results indicate the control scheme can achieve drag reduction near the training point, but a potential improvement in sensitivity to lower Reynolds numbers may be achieved with an adaptive model such as online DMD \citep{deem_2020}.
\par
\subsection{Effect of three dimensionality}\label{sec: SRS}
Clearly, the model and controller are trained on a 2D URANS flow field, which is not guaranteed to be a valid approach to what is fundamentally 3D flow due to the initiation of secondary instabilities \citep{Williamson_1996}. To validate our hypothesis that 2D URANS training data are sufficient for controller designs at $\Rey$\,$=$\,$1000$, 3D simulations of the cylinder flow field are required \citep{karniadakis_1992, mittal_1995, rodi_1998}. We assess such a design on more complex dynamics by applying the controller to a 3D simulation using the delayed-detached eddy simulation model (DDES) with uniform actuation. DDES is a modification of the detached eddy simulation treatment of turbulence, which aims to combine RANS-like behaviour in the boundary layer whilst enabling large eddy simulation resolution in the separated regions \citep{Spalart_06}. For cylinder flows, this results in laminar boundary layer separation determined by the RANS model near the wall, as in the 2D simulations. But in the wake, larger eddies are resolved, subject to the grid size, by resolving larger-scale eddies in the wake, the DDES application improves upon the accuracy of non-linear interactions available to the largest eddies \citep{spalart_2000}.
\par
\begin{figure}[h!]
\centerline{ 
\includegraphics[width=\textwidth]{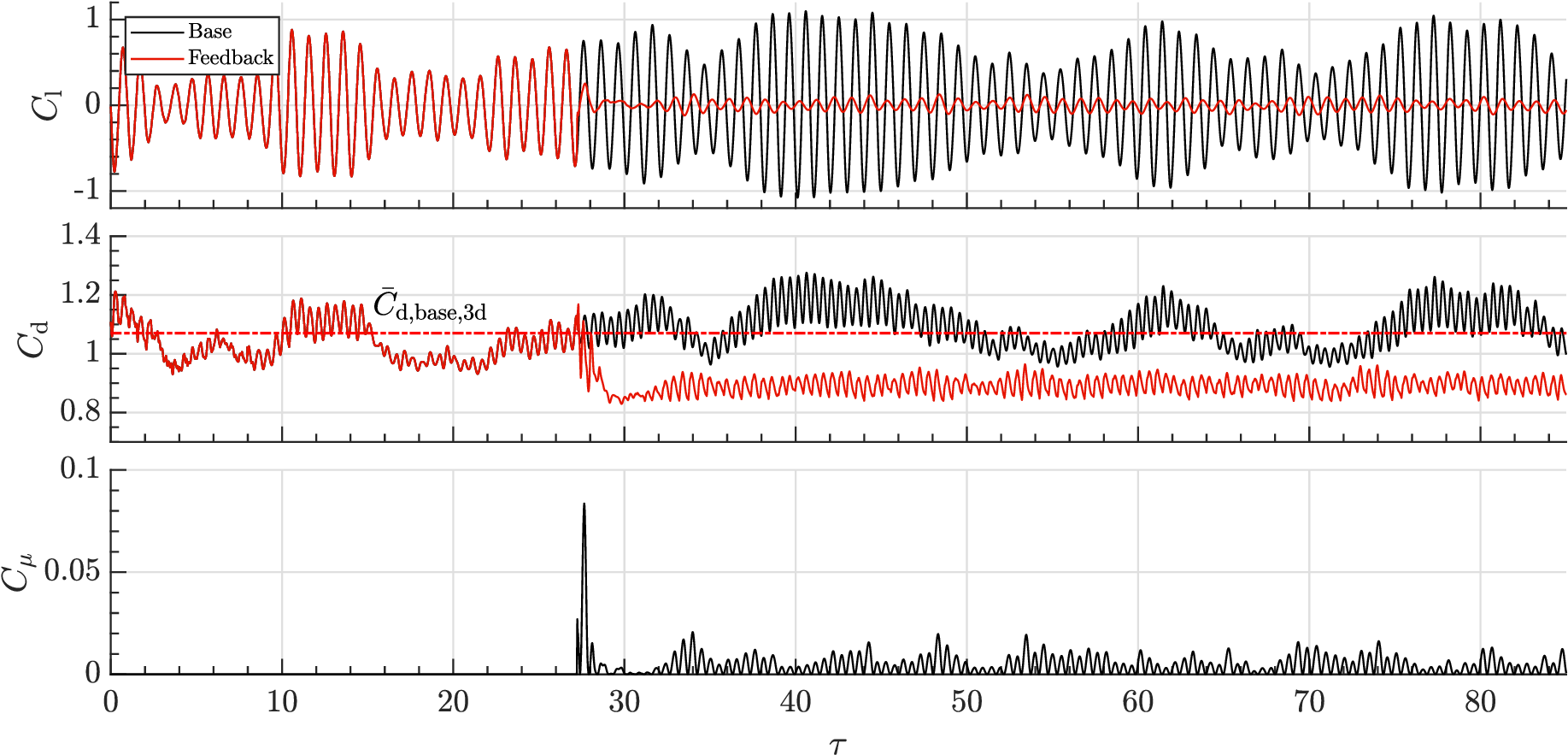}}
\caption{Lift, drag and momentum coefficient for 3D DDES results comparing baseline and controlled case.}\label{fig: srsresults}
\end{figure}

The fluid domain is extruded by a spanwise length $z$\,$=$\,$2D$ with periodicity applied at the spanwise boundaries, following the approach of \citet{travin_2000}. Notably, other studies using scale-resolving simulation for bluff bodies use a range of spanwise lengths from $z$\,$=$\,$[D,10 D]$ \citep{jordan_1998, Tong_Cheng_Zhao_2015}. It appears there is a small but acceptable difference in results for these different spanwise lengths. 30 cells subdivide the spanwise direction. The simulation is initially run for 20 shedding cycles before information is collected for a further 80 shedding cycles to ensure aerodynamic properties are statistically converged. The controller is turned on at $\tau$\,$=$\,$27$. 
The results showing lift, drag, and momentum coefficient comparing the baseline and actuated cases using the best performing $11{th}$-order controller with tuning parameters $\mathsfbi{Q}_\text{LQR}$\,$=$\,$\mathsfbi{C}^\mathrm{T}\mathsfbi{C}$, $\mathsfbi{R}_\text{LQR}$\,$=$\,$7.5\times10^{-6}$, $\mathsfbi{Q}_\text{f}$\,$=$\,$1$, $\mathsfbi{R}_\text{f}$\,$=$\,$3\times 10^{-4}$ are shown in figure \ref{fig: srsresults}.
\par
A summary of the results from the baseline simulation at $\Rey$\,$=$\,$1000$ is provided in table \ref{table:DEScompare}, along with results from other 3D high-fidelity simulations and experiments. The shedding Strouhal number agrees with higher-fidelity DNS and experimental work, and is unchanged from 2D URANS results. The mean drag coefficient is lower than 2D simulations due to the resolving of complex flow structures, resulting in spanwise variation leading to a narrower wake \citep{mittal_1995}.
\begin{table}
  \begin{center}
\def~{\hphantom{0}}
  \begin{tabular}{lccccccc}
      Case  & $L_z/D$   & $\upDelta z/D$ & $St$ & $\bar{C}_\text{d}$ & $\theta_\text{s}$ \\[3pt]
    Current study, DDES   & $2$ & $0.067$ & 0.210 & 1.07& $83^\circ$ \\
    \citet{Jiang_Cheng_2017}, DNS  & $6$ & $0.05$ & 0.211 & 1.01 & - \\
    \citet{Tong_Cheng_Zhao_2015}, DNS   & $10$ & $0.1$ & 0.215 & 1.08 & - \\
   \citet{PAPAIOANNOU_YUE_TRIANTAFYLLOU_KARNIADAKIS_2006}, DNS  & $3\upi$ & 0.146 & 0.216 & 1.03 & - \\
    \citet{Norberg_1987}, experiment  & - & - & 0.213 & 0.98 & $\approx80^\circ$ \\
    \citet[pp. 22]{schlichting_2017}, experiment$^{(a)}$  & - & - & 0.21 & $\approx1.2$ & $\approx 80^\circ$ \\
  \end{tabular}
  \caption{Summary of 3D DDES results at $\Rey$\,$=$\,$10^3$ compared to previous studies. Results in \citet[pp. 22]{schlichting_2017} are for the entire subcritical regime. \citet{Norberg_1987} is for $\Rey$\,$=$\,$3$\,$\times$\,$10^3$ flow.} \label{table:DEScompare}
  \end{center}
\end{table}
\par
The performance of the controller is degraded, achieving a power-saving of $\zeta$\,$=$\,$16\%$ due to secondary instabilities not included in the ROM, requiring the controller to react by increasing actuation effort in an attempt to damp out these secondary instabilities.
\par
This may be attributed to the SISO architecture of the controller, in which the observable is the integrated quantity $C_\text{l}$ and actuation is spanwise uniform. Any non-uniform spanwise instabilities are therefore both unobservable and uncontrollable. Despite this, $C_\text{d}$ is reduced by $13.7\%$ with an average momentum coefficient of $\bar{C}_\mu$\,$=$\,$0.0042$. This is a comparable reduction in $C_\text{d}$ to that achieved by \cite{Chatzimanolakis_24} when extending from 2D to 3D, who utilised a reinforcement learning-based controller with eight distributed actuation slots whilst using significantly more observables (force coefficients and distributed surface sensors). 
\par
\begin{figure}[h!]
\centerline{ 
\includegraphics[width=\textwidth]{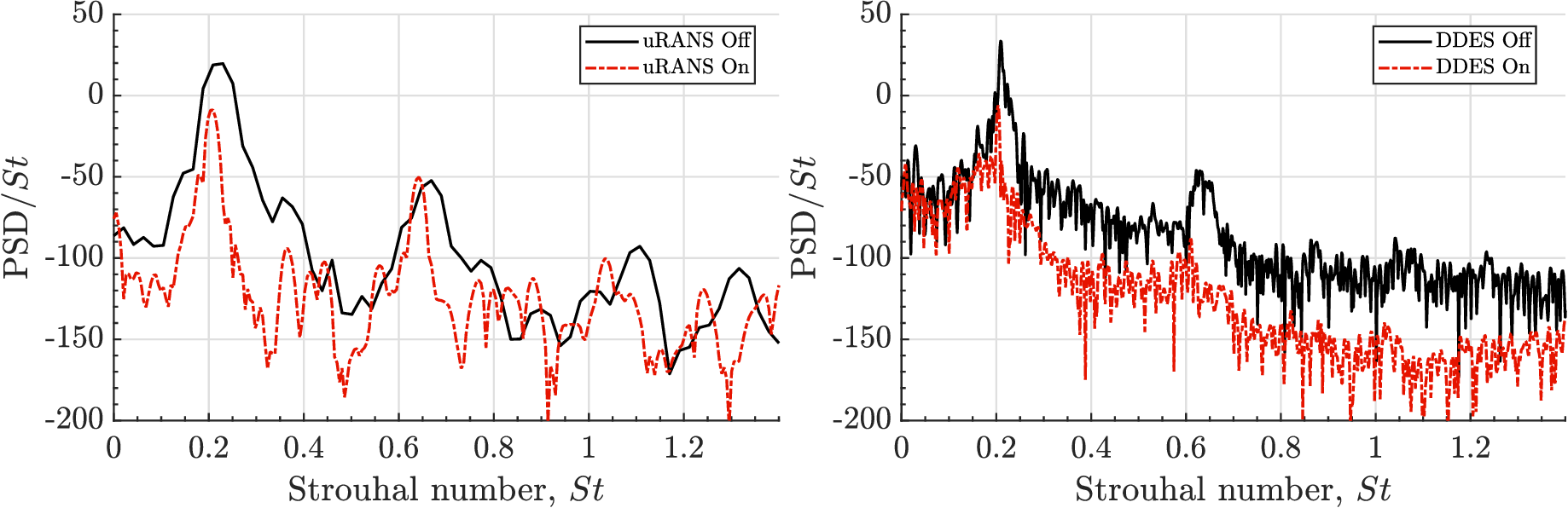}}
\caption{Power spectral density of lift coefficient comparing uncontrolled and controlled URANS and DDES simulations.}\label{fig: PSD}
\end{figure}

To verify the influence of the controller, the power spectral density of the lift history for both 2D URANS and 3D DDES results are presented in figure \ref{fig: PSD}. URANS predictions achieve a $28.6$\,dB reduction in the peak of the power spectrum associated with the lift coefficient variance, whereas, remarkably, the controller applied to DDES achieves a $39.5$\,dB reduction. The greater reduction with DDES control is due to a higher peak at $f_\text{s}$ with no control associated with increased unsteadiness due to spanwise instabilities; the controlled peak in PSD is greater than the corresponding peak with URANS. Further, the frequency of vortex shedding is reduced slightly due to the placement of closed-loop poles in the system.
\section{Conclusions}\label{sec: Conclusion}
In this study, Linear Quadratic Gaussian control derived using a reduced-order model identified from a 2D URANS simulation of flow past a cylinder at $\Rey$\,$=$\,$1000$ is demonstrated. The model is created using the dynamic mode decomposition with control algorithm. Direct measurements of the lift coefficient were used with an open-loop mapping of states to outputs created using the pseudo-inverse. By requiring the controller to drive the lift coefficient to zero, we implicitly suppress vortex shedding and reduce drag. An $11{th}$-order model was found to be suitable, capturing $77\%$ of the training data variance. Models with fewer modes were found to be substandard, with degraded performance attributed to excitation of unmodelled higher-order modes, as shown by sensitivity responses. Use of larger models leads to controller designs that suffer from the waterbed effect, in which the attempt to suppress higher-frequency unstable modes results in increased sensitivity to disturbances at frequencies outside the natural vortex shedding mode, leading to excitation of other dynamics in the flow field and non-linear interactions.
\par
Open-loop capability was tested, and a phase-shift phenomenon was accurately predicted, highlighting why feedback is necessary to suppress vortex shedding. Controller performance is calculated as the reduction in drag compared to base flow in the limit cycle, normalised by drag reduction for complete suppression of vortex shedding. The best controller achieves a performance of $\eta$\,$=$\,$75\%$, with recirculation length doubling and a 44.5\% increase in base pressure coefficient. The performance is limited by the excitation of non-linearities with higher actuation velocities. The controller achieves this with an average momentum coefficient of $\bar{C}_\mu$\,$=$\,$0.0052$. When compared to constant steady blowing at the same $\bar{C}_\mu$, feedback is shown to be a far more efficient strategy. 
\par
Subject to additional Gaussian white noise in the sensor and external disturbances of $\pm10\%$ of $U_\infty$ oscillating at $f_\text{s}/5$, the controller and Kalman state estimates remain stable and control performance is unaffected. Integral action is incorporated into the control loop with an additional state, leading to zero steady-state error between a desired reference and the observed lift coefficient. This enables a non-zero average lift coefficient to be tracked. 
\par
The controller is shown to be robust to changes in the free stream Reynolds number.
Finally, the controller trained using 2D URANS data is applied to a 3D DDES simulation. The 2D controller suppresses the spanwise-coherent vortex shedding mode when applied to the three-dimensional wake. Spanwise-varying secondary instabilities, which are unmodelled, are not controlled and result in degraded performance when compared to 2D results. Extensions to modelling and controlling 3D instabilities should be investigated further.
\par
Extensions to this work include an online model as used by \citet{deem_2020} to actively update the model and controller, resulting in a linear time-varying system. A model predictive controller may be introduced to implicitly accommodate actuation saturation and rate limits to prevent the excitation of non-linearities in the flow field. Further investigations should study the stability bounds on the application of linear models to non-linear fluid systems. Work is ongoing to identify a fully non-linear reduced-order model with which feedback linearisation may enable globally stable controllers.

\begin{bmhead}[Funding]The authors would like to gratefully acknowledge the Engineering Physical Sciences Research Council and Rolls-Royce plc (EP/S023003/1) for funding this work.
\end{bmhead}

\begin{bmhead}[Declaration of interests]The authors report no conflict of interest. \end{bmhead}

\begin{bmhead}[Data availability statement]The code and data that support the findings of this study are available at the GitHub repository: https://github.com/Jackp-1/LQG-Cylinder-Control
\end{bmhead}

\begin{bmhead}[Author ORCIDs] J. Proudfoot, https://orcid.org/0000-0002-8255-0422; C. J. Nicholls, https://orcid.org/0000-0003-2819-193X; B. M. T. Tang, https://orcid.org/0000-0001-6699-5301; M. Bacic, https://orcid.org/0000-0002-2837-7032 \end{bmhead}

\begin{bmhead}[Author contributions]\textbf{Jack Proudfoot:}Writing – original draft \& editing, Methodology, Investigation, Formal analysis, Conceptualisation. \textbf{Chris J. Nicholls}: Writing – review \& editing, Supervision. \textbf{Brian M.T. Tang}: Writing - review \& editing, Supervision. \textbf{Marko Bacic}: Writing – review \& editing, Supervision, Project administration, Methodology, Conceptualisation \end{bmhead}


\bibliographystyle{jfm}
\bibliography{jfm}
\end{document}